\documentclass[sigconf,nonacm]{acmart}
\usepackage{tabularx}
\AtBeginDocument{%
  }

\begin{document}

\title{Neural Representations as New Modes of Interactivity  
}

\author{Stanley Huang}
\email{stanleyh@mit.edu}
\orcid{1234-5678-9012}
\authornotemark[1]
\affiliation{%
  \institution{Massachusetts Institute of Technology}
  \city{Cambridge}
  \state{MA}
  \country{USA}
}

\author{Anthony Baez}
\affiliation{%
  \institution{Massachusetts Institute of Technology}
  \city{Cambridge}
  \country{MA}}
\email{acbaez@mit.edu}

\author{Catherine Yeh}
\affiliation{%
  \institution{Harvard University}
  \city{Cambridge}
  \state{MA}
  \country{USA}
}

\author{Pat Pataranutaporn}
\affiliation{%
 \institution{Massachusetts Institute of Technology}
 \city{Cambridge}
 \state{MA}
 \country{patpat@mit.edu}}

\renewcommand{\shortauthors}{Huang et al.}



\begin{abstract}
Human-AI interaction today runs almost entirely through prompts and outputs, keeping AI systems opaque to general users. We introduce \textit{Neural Representation Interaction} (NRI), a paradigm that treats the internal representations of neural networks as an interaction material people can inspect, steer, and edit. Situating NRI alongside graphical, tangible, and bio-computer interaction, we argue that the concepts and algorithms models encode internally offer \textit{representational affordances} for richer human-AI interaction. We synthesize actionable mechanistic interpretability methods and, drawing on General Morphological Analysis, construct a design space spanning \textit{Interaction}, \textit{Target}, \textit{Referent}, and \textit{Purpose}. Mapping existing interfaces onto this space reveals that most serve expert auditors, while few support lay users. We then present speculative applications in education, personal user modeling, and hiring audits, and discuss design principles, limitations, and sociotechnical risks. NRI offers HCI researchers abstractions for designing interactions that work \textit{with} AI computation rather than only through its inputs and outputs.
\end{abstract}





\begin{teaserfigure}
    \label{sec:teaser-framework}
    \centering
    \includegraphics[width=\textwidth]{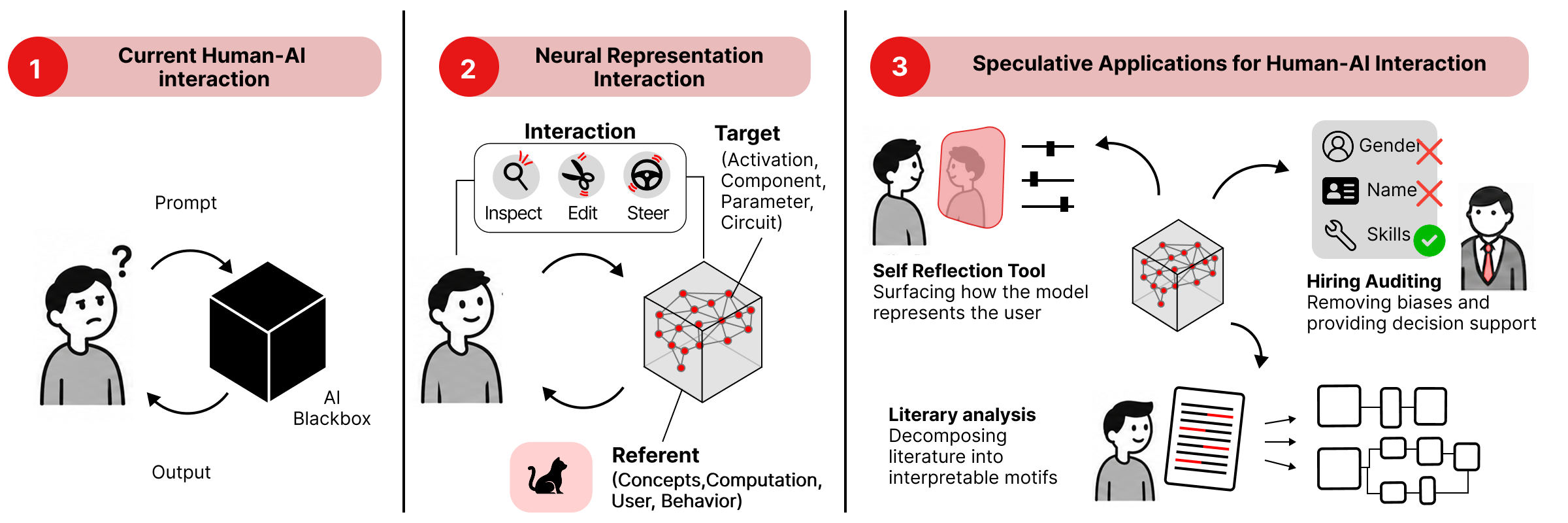}
    \caption{Mapping and conceptualizing the design space of neural representation interaction.}
    \label{fig:framework}
\end{teaserfigure}

\maketitle

\section{Introduction and Motivation}
\label{sec:intro}
Interpretability is fundamentally a human endeavor. Throughout history, understanding phenomena has given us greater agency, comprehension, and the resilience to act on them. This effort is even more critical as we are being heralded into an era of capable generative artificial intelligence (AI) systems, which are opaque 'black-boxes' and difficult to interpret. Initially dominated by the development of Large Language Models (LLMs), we are witnessing a new wave of foundation models beyond LLMs, such as vision~\cite{bai2025qwen3vl}, audio~\cite{latif2023sparks}, computer use~\cite{wang2025uitars2}, interface generation~\cite{runway2026solaris}, and world models~\cite{parkerholder2025genie3}. As AI systems continues to grow more expressive and embedded widely, the central challenge of human-AI interaction will move from accessing to comprehending these systems. Moreover, human interaction with such powerful and ubiquitous AI may invite new modes of interactions. 

Here, we aim to conceptualize a new interaction framework called \textbf{Neural Representation Interaction} (NRI). We define NRI as \textit{interaction that enables users to interpret and direct the neural representations of AI systems.} We imagine the parameter and representation space of complex AI systems to be new media of interaction that are subject to manipulation and intervention. This is in contrast to our current paradigm in human-AI interaction (HAI), prompting, where communication is done through natural language. While intuitive, this communication mode maintains the black-box nature of LLMs and may promote automation bias \cite{schemmer2022automationbias}, over-reliance \cite{bansal2021whole}, and anthropomorphism \cite{abercrombie-etal-2023-mirages}. Moreover, this interaction does not allow users to comprehend inner workings and leverage the rich affordances offered by the representational capacities of neural networks. As such, we present NRI as an alternative communicative medium of HAI: users can modify latent representations---used as internal semantic structures by AI systems---to control responses and interaction dynamics. Our chief tool in this paradigm is called mechanistic interpretability (MI), a rapidly evolving field that boasts an ambitious vision of "reverse engineering the neural network" to characterize model computation and behavior \cite{zou2023representation, bereska2024mechanistic}.

For instance, possible applications include an auditing tool for an AI resume screener, where a hiring manager can use the model's internal structure to audit whether models are displaying systemic biases by relying on names, demographic characteristics, and geographic cues. Within the NRI framework, users could scrutinize model representations and causal mechanisms for bias and make targeted edits to remove them. Accordingly, NRI enables users to act on \textit{representational bits}, or elementary units of information embedded within trained AI systems. As such, NRI is a new framework that enables new affordances, enriches user experiences, and widens the possibilities of new interaction dynamics.

This paper makes four contributions to HAI: 
\begin{itemize}
    \item First, we introduce \textit{Neural Representation Interaction} (NRI), a conceptual paradigm that treats the internal representations of neural networks as an interaction material that people can inspect, edit, and control, extending human-AI interaction beyond prompts and outputs. 
    \item Second, we provide an actionability-oriented synthesis of mechanistic interpretability methods and interfaces, connecting technical capabilities such as representation decoding, attribution, steering, and model editing to human goals including understanding, agency, decision support, and oversight. 
    \item Third, drawing on General Morphological Analysis, we develop a design space organized around four dimensions: \textit{Interaction}, \textit{Target}, \textit{Referent}, and \textit{Purpose}. We use this framework to characterize existing systems, identify underexplored combinations, and articulate progressive disclosure and iterative hypothesis testing as guiding principles for NRI interfaces. 
    \item Finally, we offer a set of speculative design propositions spanning education, personal model inspection, cognitive augmentation, and consequential decision auditing. These propositions surface opportunities, limitations, and socio-technical risks for designing interfaces that expose model internals. 
\end{itemize}


\subsection{Neural Representation Interaction as a New Framework}

\begin{table*}[t]
\centering
\caption{Interaction paradigms, their objects, and affordance principles.}
\label{tab:paradigms}
\small
\begin{tabularx}{\textwidth}{
  @{}
  p{0.24\textwidth}
  p{0.22\textwidth}
  X
  @{}
}
\toprule
\textbf{Interaction paradigm} &
\textbf{Objects} &
\textbf{Affordance principles} \\
\midrule

Graphical user interface (GUI)
\cite{sutherland1963graphical} &
Graphical objects (``painted bits'') &
Leverages visual cues, spatial reasoning, and recognition abilities. \\

\addlinespace

Tangible user interaction (TUI)
\cite{fitzmaurice1995tangiblebits} &
Physical objects (``tangible bits'') &
Leverages physical dexterity and embodied, multisensory intuitions. \\

\addlinespace

Bio-computer interaction
\cite{pataranutaporn2020livingbits} &
Microorganisms (``living bits'') &
Leverages the biodegradable and continuous properties of natural systems. \\

\addlinespace

Neural representation interaction &
Latent representations (``representational bits'') &
Leverages the predictive, generative, and representational capacities of AI systems. \\

\bottomrule
\end{tabularx}
\end{table*}

Interaction paradigms are often driven by the unique affordance principles of their respective technologies, a property of a system that suggests or enables certain actions, and motivated from limitations from earlier frameworks (see \autoref{tab:paradigms}). For instance, graphical user interfaces (GUI) \cite{sutherland1963graphical}, pioneered by Ivan Sutherland, were driven by the limitation of human recall capabilities that hindered syntax memorization for computer control. Meanwhile, tangible user interaction (TUI) \cite{fitzmaurice1995tangiblebits} aims to bridge the inconsistency of interaction on bitmapped devices with the physical world. More recently, living bits \cite{pataranutaporn2020livingbits} were conceptualized to incorporate biological interfaces and organisms to sense and actuate for more naturalized integration with the human body. Thus, we will accordingly define the affordance principles and motivations for this new framework.

First, NRI aims to leverage the unique generative and representational capacities of AI systems. We refer to the use of these capacities to augment human capacities as the \textit{computational affordances} of modern neural networks. We define generative affordances as the relation between an AI's capacity to generate novel outputs and the user's capacity to specify, shape, and evaluate those outputs. Correspondingly, representational affordances of an AI system are the coupling between the system's neural representational states and the user's capacity to perceive, interpret, and act \textit{with} the system's computation rather than through its inputs and outputs. 

Thus far, the generative affordances have been the primary focus. Notably, the affordances of deep learning have spurred rapid progress across many domains. The use of these models made significant strides in solving challenging problems and addressing user needs such as cataloging and generating new protein structures \cite{deepmind2025alphafold, liang2024foundation}, modeling global weather forecasts \cite{lam2023graphcast}, code generation, synthesizing literature review, and assisted writing. 

Yet the use of these models in a blackboxed approach has not exhausted the representational affordances. The representational capacities of models include the ability to encode concepts and algorithmic computations not observed in outputs. In the past, interactive machine learning has aimed to move users past this input-output boundary by letting users shape a model through rapid cycles of labeling, correction, and feature weighting~\cite{fails2003interactive, amershi2014power}. Feature weighting perturbs the influence of inputs but does not provide us insights to the rich patterns of a model's internal computation. An illustrative example of this is the application of AI systems to chess games, which yielded both human-identifiable \cite{mcgrath2022acquisition} and machine-level concepts \cite{schut2025bridging}. Other examples include new identifiable concepts in a well-played game of Go \cite{shin2023superhuman}, new mathematical discoveries in construction of large cap sets \cite{romeraparedes2024funsearch}, and efficient matrix multiplication algorithms \cite{fawzi2022alphatensor}. In other words, neural representations can indeed yield \textit{novel} insights for humans. Moreover, these models can encode concepts and algorithmic pathways that can be subject to direct manipulation and editing \cite{gandikota2024conceptsae, meng2022rome, rimsky2024steering}. In \autoref{sec:actionability}, we turn to an actionable review of the methods in mechanistic interpretability (MI) that enable such an approach by working with internal representations. 

\section{Background}
To describe the available tools and methods of NRI, we conduct a review of the relevant methods and techniques of MI that also lend themselves to notions of actionable interpretability \cite{orgad2026actionable}. As such, we conduct a review of MI work that could both inform insights that enrich our comprehension of models and guide actions beyond interpretability, including improved model calibration, understanding of other scientific phenomena, and enhanced user experiences. We then also survey problems and risks in the current state of HAI that NRI could stand to help address. 

\subsection{Actionable Methods in Mechanistic Interpretability} 
\label{sec:actionability}

\paragraph{Probes and steering vectors.} There is a large body of evidence that features within LLMs are represented as linear directions in the vector space created by its activation values \cite{elhage2022superposition, nanda2023emergent, zou2023representation}. Various concepts and behaviors, such as refusal \cite{arditi2024refusal}, sentiment \cite{tigges2023linear}, truth \cite{marks2023geometry}, political beliefs \cite{kim2025linear}, and spatial and temporal relationships \cite{gurnee2023language} have been shown to be effectively mediated using such linear representations. Probes can be simple linear or logistic regression models that learn these linear directions of these concepts or behaviors. Activation values have also been projected onto linear vector directions to reveal expressions of persona traits such as evil, sycophancy or hallucination \cite{chen2025persona}. Moreover, activation steering \cite{wurgaft2026manifold, turner2023steering, rimsky2024steering} has been used to push activation values in representative directions to induce desired  behavior in the model response.  

\paragraph{Verbalization methods.} Verbalization is a burgeoning direction that aim to decode neural activations into natural language. This represents a promising and scalable direction to elicit privileged or hidden communication directly from activations while subjecting opaque computations to open-ended query and inspection~\cite{li2025privileged}. Verbalization methods first extract the latent representations of a target model and provide them to an auxiliary LLM tasked with an inspection prompt. Representations are provisioned in various ways, whether patched into the auxiliary model's forward pass~\cite{ghandeharioun2024patchscopes}, concatenated into model activations~\cite{pan2024latentqa, karvonen2025activation}, or supplied directly in-context as token embeddings~\cite{chen2024selfie, frasertaliente2026nla}. Depending on the correspondence between target and auxiliary architectures, the auxiliary model may be trained on an unsupervised objective of reconstructing the activations, as in Natural Language Autoencoders~\cite{frasertaliente2026nla}; finetuned on supervised question \& answer pairs about the target's context, as in LatentQA and Activation Oracles~\cite{pan2024latentqa, karvonen2025activation}; or used training-free, as in Patchscopes and SelfIE~\cite{ghandeharioun2024patchscopes, chen2024selfie}. Currently, these techniques have been integrated into interfaces primarily used by expert human auditors for safety evaluation of target model's latent dispositions such as unverbalized evaluation awareness  ~\cite{frasertaliente2026nla} and hidden motivations ~\cite{frasertaliente2026nla, marks2025auditing}. 

\paragraph{Sparse autoencoders and dictionary learning.} Unlike supervised probing, sparse autoencoder (SAEs) and related variants are unsupervised dictionary-learning approaches that aims to decompose features into a sparse linear combination of a higher-dimensional feature space. SAEs are trained with a reconstruction objective and sparsity penalty, under the hypothesis that a sufficiently sparse code naturally decomposes activations into monosemantic, human-interpretable concepts ~\cite{elhage2022superposition}. Transcoders \cite{dunefsky2024transcoders} extend this idea from representations to computation, learning a sparse dictionary that maps a component's input to its output; crosscoders \cite{lindsey2024crosscoders}, another variant, instead learn a single dictionary shared across multiple layers or models, so that each feature carries a decoder direction at every site. These methods have been used as general purpose tools to decompose and intervene on internal representations, having been trained at scale for frontier LLMs such as GPT-4 and Claude 3 Sonnet \cite{gao2024scaling, templeton2024scaling}. SAEs have also enabled applications across multimodal processing. In auditory domains, SAE features can underpin acoustic and paralinguistic representations, used to steer models to mitigate false speech predictions, and recover human-interpretable auditory concepts \cite{aparin2026audiosae, paek2025audiolatent}. In computer vision, they supported suppression and debiasing of harmful and unwanted visual content \cite{cywinski2025saeuron, joseph2025steeringsae}, enabled precise concept editing \cite{kim2025conceptsteersparse, zhang2023addingcontrolsae}, and extraction of interpretable features in medical imaging \cite{wesp2026medical}.

\paragraph{Attribution.} Attributional interpretability aims to characterize which components or features within a model are responsible for a given behavior or output. Different families of attribution methods quantify the extent to which particular factors drive model behavior, including input features \cite{pan2026flashtrace, sundararajan2017axiomatic}, internal components \cite{bushnaq2025stochastic}, and training data \cite{chen2026mechanistic}. Context attribution has enabled human verification of model citation \cite{cohenwang2024contextcite} that surface context saliency has been used to expose gender bias in machine translation \cite{sarti2023inseq}. Internal attribution has helped enrich understanding of internal mechanisms such as planning, multi-step reasoning, and refusal \cite{lindsey2025biology}. Training-data attribution has allowed practitioners to audit which finetuning examples give rise to a behavior and modulate the learning signal accordingly during post-training \cite{bergen2026anatomy, chen2026mechanistic}. Component attribution has also further supported circuit discovery, the identification of a sparse sub-networks that faithfully recover model behaviors \cite{conmy2023towards}. 

\subsubsection{Comparison of MI with Explainable AI}

Explainable AI (XAI) \cite{dwivedi2023explainable} methods and MI share the aim to explain and make model behaviors more transparent to users. However, unlike XAI, MI is uniquely concerned with describing how the model behaves through its internal mechanisms \cite{bereska2024mechanistic}. XAI can also include blackboxed approaches, where analysis only operates on input and output values as in LIME \cite{ribeiro2016should} or SHAP \cite{lundberg2017unified}. MI has the ambitious objective of reverse-engineering the neural network's implemented algorithms into causal mechanisms and circuits. Moreover, post-hoc XAI methods have advocated for model-agnostic mechanisms and scalability of methods, which differ from MI which can yield different results for different DNNs. While, notion of "mechanistic" has gone lax \cite{saphra2024mechanistic}, moving from definitions of causal analysis to general interaction with model internals, we adopt the definition of parsimonious and faithful explanation through characterization of internal structure and computation. Both communities have used saliency maps, counterfactuals (i.e., interchange intervention), gradient-based attribution, and autoencoders, while circuit analysis and causal scrubbing remain largely confined to MI \cite{gyevnar2026bridging}. 

\subsection{Problems and Risks in Human-AI Interaction}
\label{sec:problems}

Interpretability and transparency have been theorized as ways to remedy the psychological and cognitive risks that can emerge from AI interaction. These interaction risks include over-reliance or the tendency to over-trust model responses despite accuracy \cite{10.1145/3449287, bansal2021whole}, and anthropomorphization or the ascribing of human and personal qualities to AI models \cite{10.1145/3630106.3658956, article, abercrombie-etal-2023-mirages}. Such tendencies can exacerbate vulnerability to sycophancy \cite{cheng2026sycophantic}, emotional dependence \cite{phang2025investigating, chu2025illusions}, cognitive offloading, reduce prosocial intentionality \cite{cheng2026sycophantic}, and more. By providing interfaces that expose algorithmic decision-making, users are expected to better understand and calibrate to model responses. In practice, past paradigms such as XAI--while able to affect user experience, trust, and comprehension––often negligibly affects over-reliance \cite{bansal2021whole}, automation bias \cite{vered2023explanations, schemmer2022automationbias}, task performance \cite{schemmer2022metaanalysis}, or calibrated trust \cite{lee2004trust}. 

More recently, cognitive forcing functions, or an interaction design aimed to promote Systems 2 thinking and deeper engagement with explanations, can reduce over-reliance \cite{10.1145/3449287}. Moreover, explanations that arise when initiated by users \cite{springer2019progressive, springer2020disclosure} or during incidences of user expectation violation \cite{chen2023intuition, springer2019progressive, kizilcec2016howmuchinfo} are particularly effective. Additionally, studies that use white-box-based interpretability methods, such as reporting model token probabilities could help improve the calibration gap between model and user confidence \cite{steyvers2025calibration} and improve user's mental schema of model behavior \cite{kenny2026cwnet}. Interfaces that support these interactions, if properly designed, could allow users to invest cognitive effort to form hypotheses and integrate wider evidence from the model's neural signatures. Keeping the human in the cognitive loop has been associated with improved logical discernment \cite{danry2023askme} and decision-making \cite{miller2023evaluative}. In addition, they have the potential to expand the possibilities that novel interaction design mechanisms that strengthen user comprehension, calibrate trust, and reduce over-reliance.

\section{Design Space Framework}
\label{sec:nri-framework}
\begin{figure*}[t]
    \centering
    \includegraphics[
        width=0.95\textwidth,
        keepaspectratio
    ]{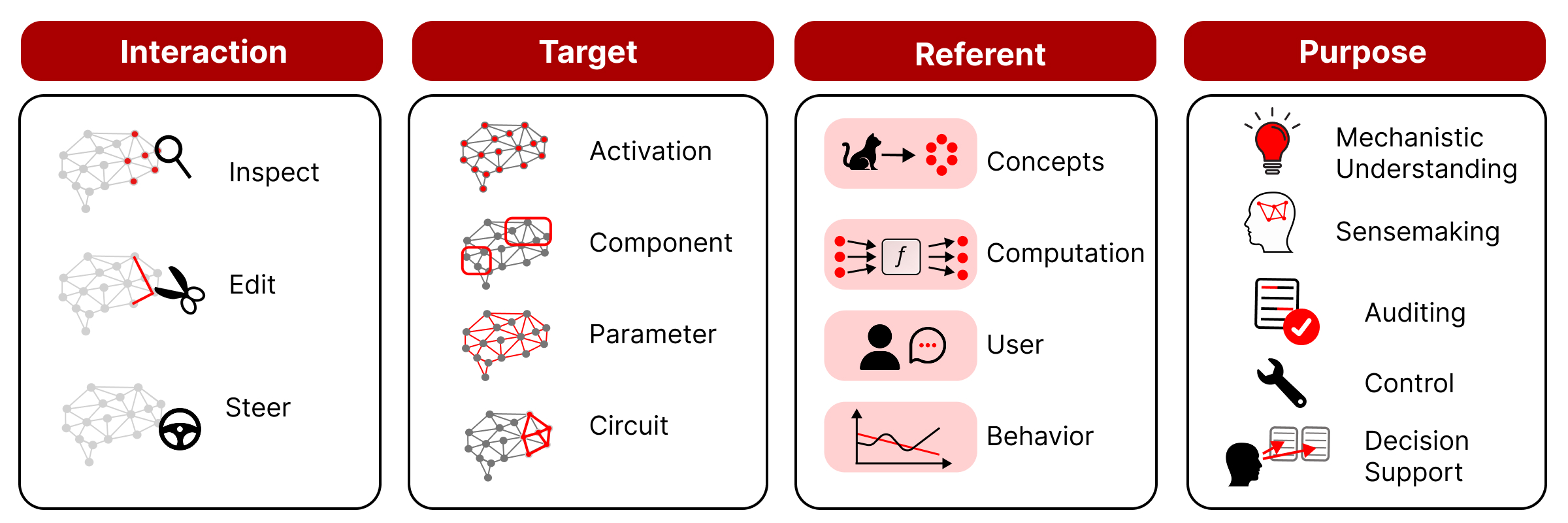}
    \caption{A multidimensional design space framework for leveraging mechanistic interpretability to enhance human-AI interactivity.}
    \label{fig:design-space}
\end{figure*}

We present a design framework to conceptualize how NRI can enable users to interact with representations or parameters of neural networks as an interaction surface. At its core, this paradigm leverages the semantic and functional richness of representational bits, presenting ways to allow users to comprehend and control them to create new human-AI experiences.

\textit{Methodology for construction of design space.} Given the vast solution space of NRI that MI and HAI could open up, we drew on General Morphological Analysis (GMA) \cite{zwicky1969discovery, ritchey2018gma} to conceptualize a forward-looking design space for current and future methods that intersect these fields. GMA is a method for reasoning about problems that are combinatorial and resist quantification by decomposing a domain into a set of facets, each admitting a discrete range of values, such that the full space of possibilities is the Cartesian product of those facets \cite{zwicky1969discovery, ritchey2018gma}. To characterize the design space, we extracted and analyzed recent surveys and scoping reviews of the MI landscape \cite{bereska2024mechanistic, rai2024practical, sharkey2025open, sawant2026mechanistic} recently published from 2024-2026. We extracted all interpretability techniques and grouped techniques by corresponding families. We, then, filtered based on the actionability principle by retaining if source literature demonstrates use beyond explanation and descriptive work (\autoref{sec:actionability}). Finally, we conducted affinity mapping to extract the defining categories and parameters, paying careful attention to the actionability principle (\autoref{sec:actionability}) and cross-consistency principle of GMA (i.e., cell products are feasibly combined) \cite{ritchey2018gma}. 

Applying GMA, we were able to decompose existing works with respect to a given morphology and then \textit{recombine} ideas to craft new analytical solutions. Our framework aims to be expansive and forward-looking by allowing ideas and methods between HAI and MI to mix and cross-pollinate to generate new ideas and applications. For instance, our review of MI from the lens of actionability revealed that much work has focused on AI safety across diverse methods tailored for technical users. Our framework enables us to also ask whether we can leverage the toolset from MI to develop tools accessible for lay users. Specifically, can we combine verbalization and multi-modal interfaces to create new affordances, interactions, and enriching user experiences? Could we provide visibility to inspect model confidence for calibrated user decision-making? Given the current emphasis of actionability on AI safety and personalization (see ~\autoref{sec:problems}), we hope our framework inspires novel interactions and synergies between different methods, applications, and visualization mediums to broader user groups. 

We describe our framework (\autoref{sec:nri-framework}), which features 4 categories of \textit{Interaction}, \textit{Target}, \textit{Referent}, and \textit{Purpose}:  

\paragraph{Interaction.} How does the user engage with the model internals? We identified three modes of interaction that enable users to interact with model internal representations:  \emph{Inspect}, \emph{Steer}, and \emph{Edit}. \emph{Inspect} allows users to read the internal representations without altering them. \emph{Edit} refers to targeted parametric or representational modifications via precise edits and ablations. \emph{Steer} enables users to induce or alter model behavior using changes in representations. Generally, steering refers to behavior on a continuous space and can be intervened based on magnitude. 

\paragraph{Target.} What internals or latents are accessed in the model? These internal features can be \emph{activations}, \emph{components}, \emph{circuits}, or \emph{parameters}. These are also the representational bits described earlier. We use \emph{activations} to broadly describe the representations of a neural network encoded in its activation values at each layer, \emph{components} as the contributions or outputs of each of its constituent architectural units. Within a transformer architecture paradigm, these may include attention heads, feedforward networks, or convolutional channels. In this sense, some models like diffusion models may not have components. \emph{Circuits} are a minimal subgraph of a neural network that implements a behavior. Finally, \emph{parameters} are the weights or biases of a neural network that encode learned knowledge from training. 

\paragraph{Referent.} What are the internals accessed taken to represent? Model representations have been understood to represent \emph{concepts}, which are specific semantic features or factual knowledge; encode algorithms or \emph{computations} that accomplish complex tasks such as indirect object identification; store knowledge or characteristics of \emph{user information} akin to theory of mind construction; and are responsible for or contribute to a specific model \emph{behavior} in output. 

\paragraph{Purpose.} What are the outcomes or desiderata of human-AI interaction we wish to aim toward? The broad categories identified are \emph{Mechanistic Understanding} as in promoting intuition over model latents and implemented algorithms; \emph{Sensemaking} that uses interpreted machine concepts or solutions to further human understanding of some phenomenon; \emph{Decision Support} to provide quantifiable or structured signals to calibrate user's confidence or decision-making; \emph{Auditing} which helps debug and surface failure modes or biases; \emph{Control} to modulate model behavior in such a way as to offer adaptive, personalized, and appropriate experiences to users. 

\paragraph{Cross-dimension considerations.} Across the five dimensions, we also emphasize two overarching facets worthy of consideration that do not fit as another dimension in our framework. The first is the \emph{end user} at hand. For instance, a technical user may want to audit a circuit to study a model behavior, whereas a non-technical user who asks for relationship advice may want to see which parts of their personal experiences induce the model to respond in a particular way (see ~\ref{sec:general-audience-application}). Progressive disclosure design methodologies and hypothesis-feedback design theories provide principled ways to provide suitable interfaces for users by considering levels of complexity and iterative cycles of work. Secondly, we should consider \emph{second-order effects} that emerge beyond static interaction. In other words, are there interaction effects and dynamics that emerge from the interplay between representational bits and user cognition? Could enabling users to programmatically interface and direct model responses lead to new understanding of AI systems? Such questions can enrich our understanding of the affordances and drawbacks to help us refine this new paradigm of interaction. Finally, applying Zwicky's cross-consistency principle, that is ensuring all condition choices between dimensions can be feasibly recombined, we note that steering is not compatible with weight space. Parameter have soft incompatibility issues with user information which is often measured with the model activations. 

\paragraph{Visualizations.} Different selections naturally give rise to different rendering mediums and ways to visually express certain features. For instance, spatial clusters or cluster-based visualizations are a natural medium to express multidimensional features in a way that preserves spatial relations. These can be applied to analysis with activations and the parameter space. Directed acyclic graphs (DAGs) can be used to represent attribution graphs and circuits by showing the cross-layer pathways of interconnected features \cite{lindsey2024crosscoders}. Grids have been used by methods like logit lens to show important discrete and qualitative concepts at different time steps \cite{lesswrongInterpretingGPT}. Other visualization methods like dendrograms could be used to visually describe symbolic relationships that compose across layers with increasing levels of abstraction \cite{mccoy2026emergent}. 

\section{Applying Design Space to Current Tools}
\label{sec:existing-works}

\begin{table*}[t]
\centering
\small
\begin{tabular*}{\textwidth}{@{\extracolsep{\fill}} l l l l l @{}}\toprule
\textbf{Interface} & \textbf{Interaction} & \textbf{Target} & \textbf{Referent} & \textbf{Purpose} \\
\midrule
GAN Dissect \cite{bau2019gandissect} & Inspect, Edit & Components & Concept & Mechanistic Understanding, Control \\
GAN Paint \cite{bau2019ganpaint} & Inspect, Edit & Components & Concept & Control \\
TalkTuner \cite{chen2024designing} & Inspect, Steer & Activations & User information & Control, Auditing \\
Neural Transparency \cite{karny2026neural} & Inspect & Activations & Behavior & Decision support \\
Multi-Turn Neural Transparency \cite{karny2026multi} & Inspect & Activations & Behavior & Decision support \\
Circuit Tracer \cite{ameisen2025circuit} & Inspect, Edit & Circuits, & Computation, Behavior & Mechanistic understanding \\
 & & Components & & \\
Assistant Axis \cite{lu2026assistantaxis} &  Edit & Activations & Behavior & Control, Auditing \\
Goodfire Ember \cite{goodfire2024ember} & Inspect, Steer & Activations & Concept, Behavior & Control, Auditing \\
ConceptViz \cite{li2025conceptviz} & Inspect, Steer & Activations & Concept & Sensemaking \\
SAE Semantic Explorer \cite{yan2025visual} & Inspect & Activations & Concept & Sensemaking \\
Concept Explorer \cite{marcilio2026navigating} & Inspect & Activations & Concept & Sensemaking \\
\bottomrule
\end{tabular*}
\caption{Examples of existing interpretability interfaces situated in the four dimensions of the NRI design space. Only TalkTuner, Neural Transparency, Multi-Turn Neural Transparency fall under interfaces for lay users.}
\label{tab:existing-works}
\end{table*}
We now turn our examination to previous, yet limited, work in intersection between MI and HCI to consider what directions might naturally arise and opportunities available. For each corresponding visualization, we apply our design space framework (Figure~\ref{fig:framework}) to showcase the general directionality most tools have been focused on.

\subsection{Interfaces for Everyday Users}
\label{sec:general-audience-application}

The primary motivations for AI use range from companionship \cite{zaosanders2025genai, zhang2026companions, miller2026unconditional, rousmaniere2025llm} and advice-seeking to domain-specific task execution \cite{chatterji2025chatgpt, handa2025economictasks}. Much work shows lay users generally prefer signals of model confidence and context \cite{ding2025citations, zhang2020confidence, bhatt2021uncertainty} since their aim is to gain calibrated trust \cite{lee2004trust} and control \cite{ehsan2024seamful} over their own interactions rather than an understanding of model internals. Notably, this involves progressively disclosing \cite{springer2020disclosure} users to traits or characteristics that may mediate model consistency, performance, and veritability. Current HCI applications have used dashboards to offer users a way to characterize their own interactions and form mental models of AI systems ~\cite{karny2026multi, karny2026neural, viegas2023system}. For instance, \textsc{Neural Transparency} presents sunburst visualizations that display predicted safety-relevant trait expressions in a custom chatbot using persona vectors derived from system prompts~\cite{karny2026neural}. A follow-up, \textsc{Multi-Turn Neural Transparency}, provides dynamic interfaces that continuously update over multi-turn conversations~\cite{karny2026multi}. These novel technologies primarily emphasize inspection of model activations to probe for personality traits that may influence model responses with a primary goal of calibrating the user's mental understanding of model behavior. Under our design framework, Neural Transparency and Multi-Turn Neural Transparency fulfill the conditions of \textit{Inspect}, \textit{Activations}, \textit{Behavior}, \textit{Decision support}. Another interactive system, \textsc{TalkTuner}, enables general users to view and directly manipulate the model's internal representation of their own demographic attributes, and was found to improve transparency and user experience~\cite{chen2024designing}. \textsc{TalkTuner}, thus, enables users to \textit{Inspect} and \textit{Steer} the linear directions of the activation space that encode \textit{User information} in order to \textit{Control} responses and \textit{Audit} how behavior may be conditioned on inherent demographic biases. 

\subsection{Interfaces for Auditors and Safety Researchers}
\label{sec:safety-applications}
To date, where MI interfaces have been designed, auditors and safety researchers are the primary beneficiaries of these novel toolings \cite{neuronpedia, ameisen2025circuit, goodfire2024ember, lieberum2024gemmascope}. Unlike interfaces for general audiences, these tools expose model internals and provision technical users with raw weights, features, activations, and circuit-level connections. Moreover, these interfaces provide greater degrees of freedom which enables more granular exploration of novel methodologies on diverse task settings and applications. These systems are often interactionally bidirectional; as such, they allow users to perform read-and-write operations \cite{ameisen2025circuit, lindsey2025biology}. Often the principle \textit{Purpose} that motivates the design of the tooling is the need for rapid hypothesis-testing feedback loop and refining researcher's raw intuition of methodologies and behavior. One example is \textsc{Circuit Tracer} which allows end users to \textit{Inspect} by tracing \textit{Circuits} in a direct graph and \textit{Edit} activations in order to see the effects on the downstream computation and behavior. While this visual modality represented internal space with nodes and edges, other tools abstract away such to provide other qualitative insights integrated into a chatbot interface. For instance, \textsc{Assistant Axis} \cite{lu2026assistantaxis} provides counterfactual views of interactive chat interfaces with and without capped activations along a helpful, "assistant" dimension. From the NRI's design framework, \textsc{Assistant Axis} enables users to \textit{Edit} a low-rank \textit{Activation} subspace that provides assistant-like mannerisms to \textit{Control} persona drift. The counterfactual views satisfy the \textit{Auditing} criterion by exposing latent biases. These tools are deployed on \textsc{Neuronpedia}, a shared infrastructure to demo MI interfaces, and so have been accordingly designed with an emphasis on providing users raw intuition of model behavior and methodological details.

\section{Speculative Applications for Designing with NRI}
NRI invites us to imagine AI internals as a new material for HAI that users can explore, manipulate, and interact with. To demonstrate the potential of NRI, we conjecture how this interaction framework could revolutionize disparate fields such as education, critical decision making, personal reflection, and AI oversight by exploring speculative applications designing for HAI with NRI. These scenarios build on emerging methods for extracting interpretable features, intervening on activations, tracing computational circuits, and testing them through causal experiments. By providing access to model internals, these applications could enable forms of inquiry and control that is difficult or impossible in current interfaces. Collectively, they envision a future in which people have greater agency to shape how AI understands the world and themselves, and to exercise that greater agency in working with the AI systems. We consider both the benefits and potential harms that such a paradigm could result in if it came to fruition.

\subsection{Mechanistic Education: Shakespeare Writing Studio}
\label{subsec:mechanism-quest}

Shakespeare's writing is celebrated for its extraordinary command of language. However, the qualities that make it remarkable also make it intimidating to study for a student new to such work. The rise of generative AI has unfortunately reduced the incentives for learning how to write in this style, since students can simply ask the models to imitate it. \textit{Deconstructing Shakespeare Studio} would apply a model's capacity for language modeling and self-explanation to this problem, letting students inspect literary topics such as the syntax, rhythm, imagery, and meaning underlying a passage's style with explanations at different levels of sophistication and nuance. Such an application could decompose the literary concept of "Shakespearean" into combinations of sparse, multidimensional concepts that represent different semantic and syntactic notions. Rather than enabling the cognitive offloading of a student's understanding, such a tool can enrich and deepen learning; the model's representation of languages can be dissected into constituent interpretable components for human learning. 

Shakespeare's style has been extensively characterized in terms of identifiable methods: metrical patterning in blank verse, including feminine endings and shared lines~\cite{wright1988metrical}; hendiadys~\cite{wright1981hendiadys}; classical rhetorical figures such as antithesis, anaphora, chiasmus, and hyperbaton~\cite{joseph1947arts,adamson2001reading,mcdonald2001arts}; functional shift, in which words migrate across grammatical categories~\cite{davis2007thinking}; and more \cite{booth1977sonnets, clemen1987soliloquies, vickers1968prose, spurgeon1935imagery}. A student examining a passage from \emph{Macbeth} could select lines and compare the constituent components that are highly activated and track how use of these different literary devices varies across samples. Students could further explore if any new abstractions are represented in the model, aiding in their own analysis. This same approach could generalize beyond Shakespeare to any written work of any genre, author, or subject: the rhetoric of celebrated speeches, the argumentative structure of noted essayists, or the harmonic and rhythmic patterns studied in music education, letting students connect any passage to broader questions or deconstruct how style and persuasion are produced for their own personalized learning. 

Under the design space framework, students could first \emph{inspect} passage-specific \emph{activations}, then \emph{steer} models at different points in the input sequence to see how texts alter due to earlier changes. Allowing users to understand the stylistic fluency in this way, in terms of first principles, is an example of literary \emph{Sensemaking}. Moreover, this example illustrates the sensemaking process and places the emphasis on decoding and reinterpreting texts with the underlying motifs and literary building blocks.

\subsection{Transparent User Models: Latent Mirror}
\label{subsec:latent-mirror}

One step beyond current user modeling done by LLMs would be a \textit{Transparent User Model}. Advances in AI systems and HCI techniques could lead to models that can learn your habits, preferences, and identity in personalized ways from multiple different modalities of data, such as phone and computer usage, internet of things, and AI assistant interaction. Simultaneously, advances in MI could enable us to interpret the representations models use in their hyper-personalized and aware user model. Such user models could lead to interfaces that allow users to query their own models of themselves: users could ask models what the system believes to be recurring behaviors or decision-making, grounded in traceable internal representations built from months or years of accumulated interactions \cite{zhang2024personalization, liu2025surveypersonalization}.

We imagine Latent Mirror, an application that radically extends work like \textit{TalkTuner} (see ~\autoref{sec:existing-works}). Latent Mirror would allow users to inspect, attribute, and modify representations of themselves within the model's computations. For instance, a user might begin with a concern such as, ``Why does the assistant keep steering me toward low-sodium recipes and shorter walking routes?'' after noticing the pattern across several unrelated conversations. Latent Mirror would retrieve relevant portions of the interaction history and identify candidate internal features associated with this recurring response pattern. The application might propose the interpretation of an inferred chronic health condition, accompanied by the excerpts that led the model to build such an implicit profile---here, a symptom mentioned once in passing and never raised again.

Correspondingly, Latent Mirror could allow the model to programmatically \textit{steer} by capping the model's \textit{activations} \cite{lu2026assistantaxis} or to inhibit the model's assumption of a user behavior. Interventions in model responses that reduce persona drifts or alter the model's implicit profile of the \textit{user} could affect downstream responses. The user could test its functional role by suppressing the responses and comparing the differences in responses prior to intervention. In terms of the design space, Latent Mirror would support \emph{inspection} and even \emph{steering} or persistent \emph{editing}. Its primary \emph{targets} would be contextual \emph{activations} that encode certain identify-relevant \emph{concepts}. These targets would refer to inferred \emph{user information}, including preferences, goals, emotional cues, and interaction patterns, as well as the model \emph{behavior} influenced by those inferences. The resulting interaction would demonstrate \emph{control}, as they alter the user model that is formed. Modifications of models' belief state could also be used to support more supportive and healthier human-AI interaction dynamics, such as extending to use cases like adaptive tutoring or accessibility.

\subsection{Causal Auditor: Auditing Bias in Hiring}
\label{subsec:equity-lens}

AI-assisted hiring systems can influence how recruiters interpret résumés and allocate attention across a candidate pool \cite{wilson2024gender, anzenberg2025evaluating}. A recommendation may appear reasonable because it comes with a fluent explanation, yet that explanation may not properly attribute the computation or reasoning that produced it. For instance, a model might cite limited leadership potential while actually making its decision based on institutional prestige, career gaps, or names that serve as demographic proxies. Conversely, a model detecting demographic information internally does not mean that information causally impacted the recommendation. This gap makes it difficult for recruiters or auditors to determine when an assessment deserves scrutiny.

Causal Auditor is a speculative application that would help resume reviewers and auditors investigate such factors in opaque hiring decisions, enabling the investigation of testable hypotheses about how the model used the evidence at hand to make its determination. NRI breaks this opacity, allowing users to scrutinize representations and circuits to aid in ensuring decisions are made equitably, fairly, and effectively, targeting representations responsible for bias similar to prior work \cite{vig2020investigating}. For instance, a reviewer might select an outputted model justification such as ``The candidate demonstrates limited leadership potential.'' The interface would attribute this sequence to résumé excerpt that influenced it and surface internal features tied to management experience, institutional prestige, or employment gaps. Each of these would be backed by examples and counterexamples from an evaluation corpus \cite{pena2020bias}, and counterfactuals could be tested to more rigorously establish causality. The reviewer could compare two matched résumés that differ only in a suspected proxy, such as an institutional name, and Equity Lens would evaluate where their internal pathways diverge. Swapping concepts between the two traces and observing whether the assessment shifts tests that pathway causally \cite{meng2022locating}. This would mechanistically and faithfully diagnose whether systemic biases exist in such models, ensuring transparency in previously opaque systems. These models could then be steered or these biased representations suppressed in activation to move closer to an unbiased decision, mirroring previously done interventions \cite{zou2023representation}.

Within the design space, Equity Lens would enable reviewers to \emph{inspect} internal representations and \emph{edit} them within a controlled audit environment. Its \emph{targets} would include \emph{activations} tied to application evidence, \emph{components}, and \emph{circuits} that link sensitive conceptual referents such as gender, name, race, or biographical information to output. This supports \emph{auditing} and \emph{decision support}: understanding how an assessment was produced, and using that to calibrate reliance or revise decisions. The system would keep an audit record of the concern, targets, interpretations, interventions, and alternative explanations, usable for independent review. Similar approaches could extend to admissions, credit, and benefits review, anywhere outcomes may need to be properly adjudicated to avoid use of protected characteristics. 

\section{Limitations}
\label{sec:limitation}

We present NRI as a forward-looking paradigm, and its limitations are correspondingly both technical and socio-technical. There are fundamental assumptions such as model architecture, maturity of underlying methods, access provisioning, and how current interfaces present representational bits. 

\paragraph{Architectural assumptions of the design space.} Our design space, and much of the MI literature it synthesizes, is grounded in transformer language models. The \textit{Target} facet in particular presumes a residual stream, discrete attention and feedforward components, and circuits as subgraphs over them. Transferring such architectural prescriptions to other architectures could not be possible. Steering, latent-prediction lenses, and factual editing have been shown to carry over to state space and recurrent models such as Mamba and RWKV, though component-level analysis must be reformulated around a compressed state rather than attention heads~\cite{paulo2025rnn, sharma2024mamba}. In diffusion models, dictionary learning has enabled concept suppression and editing~\cite{cywinski2025saeuron, kim2025conceptsteersparse}, but there is no comparative analogy for a circuit. For world models~\cite{parkerholder2025genie3}, interpretability tooling is largely absent. As such, the cross-consistency of our facets is established for transformers but may not hold for others; some cells of the design space may not be comparable or require redefinition for non-transformer systems.

\paragraph{Methodological progress and reliability.} The methods reviewed in \autoref{sec:actionability} are actively being developed and improved. For instance, probes and steering vectors rest on the linear representation hypothesis, yet not all features seem to be linear~\cite{engels2024notlinear}, motivating ongoing work on manifold-based steering~\cite{wurgaft2026manifold}; existing steering vectors generalize inconsistently beyond the distribution they were derived from and produce unintended side effects on unrelated behavior~\cite{tan2024steering}. SAEs may not recover ground-truth and human-interpretable features as concepts can split across latents as dictionary size grows~\cite{leask2025canonical}. Verbalization requires an auxiliary module trained or prompted to describe activations and may risk confabulating descriptions that the target's computation does not support~\cite{ghandeharioun2024patchscopes, pan2024latentqa}. Many open technical and reliability problems of MI constrain the feasibility of what current NRI interfaces may look like. 

\section{Sociotechnical risks of NRI. }
\paragraph{Data provenance and personalization} Our speculative propositions, most directly Transparent User Models and Latent Mirror (\autoref{subsec:latent-mirror}), raise questions of data provenance that current privacy policies may not adequately address. Representations built from accumulated interactions can encode inferences the user never stated, and LLMs already infer sensitive attributes such as location, income, and demographics from innocuous text~\cite{staab2024beyond}. The ability for personalized support carries a dual-use risk, since this also could enable model providers or other actors to conduct precise tailoring of messages~\cite{salvi2025persuasion} and could be used for targeted advertisement. NRI interfaces operating on user information should adhere to the rights of access, rectification, erasure, and data portability. Future work should develop privacy and ethical frameworks specific to representational bits.

\paragraph{Anthropomorphization.} How representational bits are interfaced with may also promote the anthropomorphization which NRI aims to counter (\autoref{sec:problems}). Verbalization could convey activations in the grammar of mental states as reporting what a model "believes" or "wants" invites users to treat statistical structure as intention~\cite{shanahan2024talking, abercrombie-etal-2023-mirages}. These novel interfaces have an opportunity to be careful with their communication by describing internals in the language of \emph{algorithm} and \emph{computation}. Additionally, the interaction methods, such as editing, could be presented in a way that portrays a sense of determinism. Whether such framings do in fact reduce anthropomorphism is a potential empirical question.

\section{Conclusion}
\label{sec:conclusion}
We introduce NRI as a novel human-AI interaction framework. Our paper draws on affordance theories and past interaction paradigms to contextualize this new mode of interactivity. In addition to our design space, we also provided speculative examples and mapped existing work to our framework. As such, NRI aims to present a forward-looking and opportunistic vision: what are different ways we may interact with AI systems? How are our experiences and interpretations coupled with representational bits? How can we design systems that take advantage of such representational affordances? This paper is an attempt to introduce these new conceptual questions and provide HCI and HAI researchers with abstractions to think in novel modes of interactivity. Together, our study aims to augment human-AI interaction by transforming AI systems from inscrutable algorithms to transparent partners.




\bibliographystyle{ACM-Reference-Format}
\bibliography{reference}

@String{Computing = "Computing" }

@String{Computer = "{IEEE} Computer" }

@String{Macmillan = "Macmillan" }

@String{Springer = "Springer-Verlag" }

@article{nanda2023emergent,
  title={Emergent linear representations in world models of self-supervised sequence models},
  author={Nanda, Neel and Lee, Andrew and Wattenberg, Martin},
  journal={arXiv preprint arXiv:2309.00941},
  year={2023}
}

@article{arditi2024refusal,
  title={Refusal in language models is mediated by a single direction},
  author={Arditi, Andy and Obeso, Oscar and Syed, Aaquib and Paleka, Daniel and Panickssery, Nina and Gurnee, Wes and Nanda, Neel},
  journal={Advances in Neural Information Processing Systems},
  volume={37},
  pages={136037--136083},
  year={2024}
}

@article{tigges2023linear,
  title={Linear representations of sentiment in large language models},
  author={Tigges, Curt and Hollinsworth, Oskar John and Geiger, Atticus and Nanda, Neel},
  journal={arXiv preprint arXiv:2310.15154},
  year={2023}
}

@article{chen2025persona,
  title={Persona vectors: Monitoring and controlling character traits in language models},
  author={Chen, Runjin and Arditi, Andy and Sleight, Henry and Evans, Owain and Lindsey, Jack},
  journal={arXiv preprint arXiv:2507.21509},
  year={2025}
}

@article{dunefsky2024transcoders,
  title={Transcoders find interpretable llm feature circuits},
  author={Dunefsky, Jacob and Chlenski, Philippe and Nanda, Neel},
  journal={Advances in Neural Information Processing Systems},
  volume={37},
  pages={24375--24410},
  year={2024}
}

@article{chen2024designing,
  title={Designing a dashboard for transparency and control of conversational AI},
  author={Chen, Yida and Wu, Aoyu and DePodesta, Trevor and Yeh, Catherine and Li, Kenneth and Marin, Nicholas Castillo and Patel, Oam and Riecke, Jan and Raval, Shivam and Seow, Olivia and others},
  journal={arXiv preprint arXiv:2406.07882},
  year={2024}
}

@article{phang2025investigating,
  title={Investigating affective use and emotional well-being on ChatGPT},
  author={Phang, Jason and Lampe, Michael and Ahmad, Lama and Agarwal, Sandhini and Fang, Cathy Mengying and Liu, Auren R and Danry, Valdemar and Lee, Eunhae and Chan, Samantha WT and Pataranutaporn, Pat and others},
  journal={arXiv preprint arXiv:2504.03888},
  year={2025}
}

@misc{neuronpedia,
    title = {Neuronpedia: Interactive Reference and Tooling for Analyzing Neural Networks},
    year = {2023},
    note = {Software available from neuronpedia.org},
    url = {https://www.neuronpedia.org},
    author = {Lin, Johnny}
}

@article{templeton2024scaling,
       title={Scaling Monosemanticity: Extracting Interpretable Features from Claude 3 Sonnet},
       author={Templeton, Adly and Conerly, Tom and Marcus, Jonathan and Lindsey, Jack and Bricken, Trenton and Chen, Brian and Pearce, Adam and Citro, Craig and Ameisen, Emmanuel and Jones, Andy and Cunningham, Hoagy and Turner, Nicholas L and McDougall, Callum and MacDiarmid, Monte and Freeman, C. Daniel and Sumers, Theodore R. and Rees, Edward and Batson, Joshua and Jermyn, Adam and Carter, Shan and Olah, Chris and Henighan, Tom},
       year={2024},
       journal={Transformer Circuits Thread},
       url={https://transformer-circuits.pub/2024/scaling-monosemanticity/index.html}
    }

@article{gao2024scaling,
  title={Scaling and evaluating sparse autoencoders},
  author={Gao, Leo and la Tour, Tom Dupr{\'e} and Tillman, Henk and Goh, Gabriel and Troll, Rajan and Radford, Alec and Sutskever, Ilya and Leike, Jan and Wu, Jeffrey},
  journal={arXiv preprint arXiv:2406.04093},
  year={2024}
}

@article{gurnee2023language,
  title={Language models represent space and time},
  author={Gurnee, Wes and Tegmark, Max},
  journal={arXiv preprint arXiv:2310.02207},
  year={2023}
}

@article{elhage2022superposition,
   title={Toy Models of Superposition},
   author={Elhage, Nelson and Hume, Tristan and Olsson, Catherine and Schiefer, Nicholas and Henighan, Tom and Kravec, Shauna and Hatfield-Dodds, Zac and Lasenby, Robert and Drain, Dawn and Chen, Carol and Grosse, Roger and McCandlish, Sam and Kaplan, Jared and Amodei, Dario and Wattenberg, Martin and Olah, Christopher},
   year={2022},
   journal={Transformer Circuits Thread},
   url={https://transformer-circuits.pub/2022/toy_model/index.html}
}

@book{wright1988metrical,
  author    = {Wright, George T.},
  title     = {Shakespeare's Metrical Art},
  publisher = {University of California Press},
  address   = {Berkeley},
  year      = {1988}
}

@article{wright1981hendiadys,
  author  = {Wright, George T.},
  title   = {Hendiadys and \emph{Hamlet}},
  journal = {PMLA},
  volume  = {96},
  number  = {2},
  pages   = {168--193},
  year    = {1981}
}

@book{joseph1947arts,
  author    = {Joseph, Sister Miriam},
  title     = {Shakespeare's Use of the Arts of Language},
  publisher = {Columbia University Press},
  address   = {New York},
  year      = {1947}
}

@book{adamson2001reading,
  editor    = {Adamson, Sylvia and Hunter, Lynette and Magnusson, Lynne and Thompson, Ann and Wales, Katie},
  title     = {Reading Shakespeare's Dramatic Language: A Guide},
  publisher = {Arden Shakespeare},
  address   = {London},
  year      = {2001}
}

@book{mcdonald2001arts,
  author    = {McDonald, Russ},
  title     = {Shakespeare and the Arts of Language},
  publisher = {Oxford University Press},
  address   = {Oxford},
  year      = {2001}
}

@article{zhang2024personalization,
  title   = {Personalization of Large Language Models: A Survey},
  author  = {Zhang, Zhehao and Rossi, Ryan A. and Kveton, Branislav and Shao, Yijia and Yang, Diyi and Zamani, Hamed and Dernoncourt, Franck and Barrow, Joe and Yu, Tong and Kim, Sungchul and Zhang, Ruiyi and Gu, Jiuxiang and Derr, Tyler and Chen, Hongjie and Wu, Junda and Chen, Xiang and Wang, Zichao and Mitra, Subrata and Lipka, Nedim and Ahmed, Nesreen and Wang, Yu},
  journal = {arXiv preprint arXiv:2411.00027},
  year    = {2024},
  url     = {https://arxiv.org/abs/2411.00027}
}

@article{liu2025surveypersonalization,
  title   = {A Survey of Personalized Large Language Models: Progress and Future Directions},
  author  = {Liu, Jiahong and Qiu, Zexuan and Li, Zhongyang and Dai, Quanyu and Yu, Wenhao and Zhu, Jieming and Hu, Minda and Yang, Menglin and Chua, Tat-Seng and King, Irwin},
  journal = {arXiv preprint arXiv:2502.11528},
  year    = {2025},
  url     = {https://arxiv.org/abs/2502.11528}
}

@inproceedings{paulo2025rnn,
  title={Do Transformer Interpretability Methods Transfer to {RNN}s?},
  author={Paulo, Gon{\c{c}}alo and Marshall, Thomas and Belrose, Nora},
  booktitle={Proceedings of the AAAI Conference on Artificial Intelligence},
  volume={39}, number={26}, pages={27565--27572}, year={2025}
}

@article{sharma2024mamba,
  title={Locating and Editing Factual Associations in {M}amba},
  author={Sharma, Arnab Sen and Atkinson, David and Bau, David},
  journal={arXiv preprint arXiv:2404.03646}, year={2024}
}

@inproceedings{engels2024notlinear,
  title={Not All Language Model Features Are Linear},
  author={Engels, Joshua and Michaud, Eric J. and Liao, Isaac and Gurnee, Wes and Tegmark, Max},
  booktitle={International Conference on Learning Representations}, year={2025}
}

@article{tan2024steering,
  title={Analysing the Generalisation and Reliability of Steering Vectors},
  author={Tan, Daniel and Chanin, David and Lynch, Aengus and Kanoulas, Dimitrios and Paige, Brooks and Garriga-Alonso, Adri{\`a} and Kirk, Robert},
  journal={Advances in Neural Information Processing Systems}, volume={37}, year={2024}
}

@inproceedings{leask2025canonical,
  title={Sparse Autoencoders Do Not Find Canonical Units of Analysis},
  author={Leask, Patrick and Bussmann, Bart and Pearce, Michael and Bloom, Joseph and Tigges, Curt and Al Moubayed, Noura and Sharkey, Lee and Nanda, Neel},
  booktitle={International Conference on Learning Representations}, year={2025}
}

@inproceedings{staab2024beyond,
  title={Beyond Memorization: Violating Privacy via Inference with Large Language Models},
  author={Staab, Robin and Vero, Mark and Balunovi{\'c}, Mislav and Vechev, Martin},
  booktitle={International Conference on Learning Representations}, year={2024}
}

@article{salvi2025persuasion,
  title={On the Conversational Persuasiveness of {GPT-4}},
  author={Salvi, Francesco and Horta Ribeiro, Manoel and Gallotti, Riccardo and West, Robert},
  journal={Nature Human Behaviour}, volume={9}, number={8}, pages={1645--1653}, year={2025},
  doi={10.1038/s41562-025-02194-6}
}

@inproceedings{fails2003interactive,
  title={Interactive machine learning},
  author={Fails, Jerry Alan and Olsen Jr, Dan R},
  booktitle={Proceedings of the 8th international conference on Intelligent user interfaces},
  pages={39--45},
  year={2003}
}

@article{amershi2014power,
  title   = {Power to the People: The Role of Humans in Interactive Machine Learning},
  author  = {Amershi, Saleema and Cakmak, Maya and Knox, W. Bradley and Kulesza, Todd},
  journal = {AI Magazine},
  volume  = {35},
  number  = {4},
  pages   = {105--120},
  year    = {2014},
  doi     = {10.1609/aimag.v35i4.2513}
}

@inproceedings{bau2019gandissect,
  title     = {{GAN} Dissection: Visualizing and Understanding Generative Adversarial Networks},
  author    = {Bau, David and Zhu, Jun-Yan and Strobelt, Hendrik and Zhou, Bolei and Tenenbaum, Joshua B. and Freeman, William T. and Torralba, Antonio},
  booktitle = {International Conference on Learning Representations (ICLR)},
  year      = {2019},
  url       = {https://openreview.net/forum?id=Hyg_X2C5FX},
  eprint    = {1811.10597},
  archivePrefix = {arXiv}
}

@article{bau2019ganpaint,
  title     = {Semantic Photo Manipulation with a Generative Image Prior},
  author    = {Bau, David and Strobelt, Hendrik and Peebles, William and Wulff, Jonas and Zhou, Bolei and Zhu, Jun-Yan and Torralba, Antonio},
  journal   = {ACM Transactions on Graphics (Proc. SIGGRAPH)},
  volume    = {38},
  number    = {4},
  articleno = {59},
  year      = {2019},
  doi       = {10.1145/3306346.3323023}
}

@article{shanahan2024talking,
  title={Talking About Large Language Models},
  author={Shanahan, Murray},
  journal={Communications of the ACM}, volume={67}, number={2}, pages={68--79}, year={2024}
}

@book{davis2007thinking,
  author    = {Davis, Philip},
  title     = {Shakespeare Thinking},
  publisher = {Continuum},
  address   = {London},
  year      = {2007}
}

@book{booth1977sonnets,
  editor    = {Booth, Stephen},
  title     = {Shakespeare's Sonnets},
  publisher = {Yale University Press},
  address   = {New Haven},
  year      = {1977}
}

@book{clemen1987soliloquies,
  author     = {Clemen, Wolfgang},
  title      = {Shakespeare's Soliloquies},
  translator = {Scott Stokes, Charity},
  publisher  = {Methuen},
  address    = {London},
  year       = {1987}
}

@book{vickers1968prose,
  author    = {Vickers, Brian},
  title     = {The Artistry of Shakespeare's Prose},
  publisher = {Methuen},
  address   = {London},
  year      = {1968}
}

@book{spurgeon1935imagery,
  author    = {Spurgeon, Caroline F. E.},
  title     = {Shakespeare's Imagery and What It Tells Us},
  publisher = {Cambridge University Press},
  address   = {Cambridge},
  year      = {1935}
}

@inproceedings{10.1145/3630106.3658956,
author = {Maeda, Takuya and Quan-Haase, Anabel},
title = {When Human-AI Interactions Become Parasocial: Agency and Anthropomorphism in Affective Design},
year = {2024},
isbn = {9798400704505},
publisher = {Association for Computing Machinery},
address = {New York, NY, USA},
url = {https://doi.org/10.1145/3630106.3658956},
doi = {10.1145/3630106.3658956},
booktitle = {Proceedings of the 2024 ACM Conference on Fairness, Accountability, and Transparency},
pages = {1068–1077},
numpages = {10},
location = {Rio de Janeiro, Brazil},
series = {FAccT '24}
}

@article{article,
author = {Gomes, Sofia and Lopes, João and Nogueira, Elisabete},
year = {2025},
month = {01},
pages = {},
title = {Anthropomorphism in artificial intelligence: a game-changer for brand marketing},
volume = {11},
journal = {Future Business Journal},
doi = {10.1186/s43093-025-00423-y}
}

@inproceedings{sundararajan2017axiomatic,
  title     = {Axiomatic Attribution for Deep Networks},
  author    = {Sundararajan, Mukund and Taly, Ankur and Yan, Qiqi},
  booktitle = {Proceedings of the 34th International Conference on Machine Learning (ICML)},
  series    = {PMLR},
  volume    = {70},
  pages     = {3319--3328},
  year      = {2017},
  eprint    = {1703.01365},
  archivePrefix = {arXiv},
  primaryClass  = {cs.LG}
}

@inproceedings{pan2026flashtrace,
  title     = {Towards Long-Horizon Interpretability: Efficient and Faithful Multi-Token Attribution for Reasoning LLMs},
  author    = {Pan, Wenbo and Liu, Zhichao and Wang, Xianlong and Yu, Haining and Jia, Xiaohua},
  booktitle = {Proceedings of the 43rd International Conference on Machine Learning (ICML)},
  series    = {PMLR},
  volume    = {306},
  year      = {2026},
  eprint    = {2602.01914},
  archivePrefix = {arXiv},
  primaryClass  = {cs.LG}
}

@article{orgad2026actionable,
  title         = {Interpretability Can Be Actionable},
  author        = {Orgad, Hadas and Barez, Fazl and Haklay, Tal and Lee, Isabelle and
                   Mosbach, Marius and Reusch, Anja and Saphra, Naomi and
                   Wallace, Byron C. and Wiegreffe, Sarah and Wong, Eric and
                   Tenney, Ian and Geva, Mor},
  journal       = {arXiv preprint arXiv:2605.11161},
  year          = {2026},
  eprint        = {2605.11161},
  archivePrefix = {arXiv},
  primaryClass  = {cs.LG}
}

@misc{chen2026mechanistic,
  title         = {Mechanistic Data Attribution: Tracing the Training Origins of Interpretable LLM Units},
  author        = {Chen, Jianhui and Luo, Yuzhang and Pan, Liangming},
  year          = {2026},
  eprint        = {2601.21996},
  archivePrefix = {arXiv},
  primaryClass  = {cs.CL}
}

@misc{bushnaq2025stochastic,
  title         = {Stochastic Parameter Decomposition},
  author        = {Bushnaq, Lucius and Braun, Dan and Sharkey, Lee},
  year          = {2025},
  eprint        = {2506.20790},
  archivePrefix = {arXiv},
  primaryClass  = {cs.LG}
}

@article{conmy2023towards,
  title={Towards automated circuit discovery for mechanistic interpretability},
  author={Conmy, Arthur and Mavor-Parker, Augustine and Lynch, Aengus and Heimersheim, Stefan and Garriga-Alonso, Adri{\`a}},
  journal={Advances in Neural Information Processing Systems},
  volume={36},
  pages={16318--16352},
  year={2023}
}

@techreport{chatterji2025chatgpt,
  title       = {How People Use {ChatGPT}},
  author      = {Chatterji, Aaron and Cunningham, Tom and Deming, David J.
                 and Hitzig, Zo{\"e} and Ong, Christopher and Shan, Carl Yan
                 and Wadman, Kevin},
  institution = {National Bureau of Economic Research},
  type        = {Working Paper},
  number      = {34255},
  year        = {2025},
  month       = sep,
  doi         = {10.3386/w34255},
  url         = {https://www.nber.org/papers/w34255},
  note        = {Introduces the ``Asking, Doing, Expressing'' taxonomy;
                 Asking (advisory) ~49\% and Doing (task) ~40\% of messages}
}

@misc{handa2025economictasks,
  title         = {Which Economic Tasks are Performed with {AI}? Evidence
                   from Millions of {Claude} Conversations},
  author        = {Handa, Kunal and Tamkin, Alex and McCain, Miles and
                   Huang, Saffron and Durmus, Esin and Heck, Sarah and
                   Mueller, Jared and Hong, Jerry and Ritchie, Stuart and
                   Belonax, Tim and Troy, Kevin K. and Amodei, Dario and
                   Kaplan, Jared and Clark, Jack and Ganguli, Deep},
  year          = {2025},
  eprint        = {2503.04761},
  archivePrefix = {arXiv},
  primaryClass  = {cs.CY},
  url           = {https://arxiv.org/abs/2503.04761},
  note          = {Anthropic Economic Index; maps Claude usage onto
                   O*NET occupational tasks (domain-specific execution)}
}

@inproceedings{zhang2020confidence,
  title     = {Effect of Confidence and Explanation on Accuracy and Trust
               Calibration in {AI}-Assisted Decision Making},
  author    = {Zhang, Yunfeng and Liao, Q. Vera and Bellamy, Rachel K. E.},
  booktitle = {Proceedings of the 2020 Conference on Fairness,
               Accountability, and Transparency (FAT*)},
  pages     = {295--305},
  year      = {2020},
  doi       = {10.1145/3351095.3372852},
  url       = {https://arxiv.org/abs/2001.02114},
  note      = {Displaying calibrated confidence improves trust calibration}
}

@misc{goodfire2024ember,
  title        = {Goodfire Ember: Scaling Interpretability for Frontier
                  Model Alignment},
  author       = {{Goodfire}},
  howpublished = {Goodfire blog},
  year         = {2024},
  month        = dec,
  day          = {23},
  url          = {https://www.goodfire.ai/blog/announcing-goodfire-ember},
}

@phdthesis{sutherland1963graphical,
  author = {Sutherland, Ivan E.},
  title  = {Sketchpad: A Man-Machine Graphical Communication System},
  school = {Massachusetts Institute of Technology},
  year   = {1963},
  address = {Cambridge, MA, USA}
}

@article{sawant2026mechanistic,
  title   = {Mechanistic Interpretability for Neural Networks: Circuits, Sparse Features and Symbolic Reasoning},
  author  = {Sawant, Pranav Milind and Krej{\v{c}}{\'i}, Jakub},
  journal = {arXiv preprint arXiv:2607.07316},
  year    = {2026},
  note    = {Scoping review conducted per PRISMA-ScR},
  url     = {https://arxiv.org/abs/2607.07316}
}

@article{rai2024practical,
  title   = {A Practical Review of Mechanistic Interpretability for Transformer-Based Language Models},
  author  = {Rai, Daking and Zhou, Yilun and Feng, Shi and Saparov, Abulhair and Yao, Ziyu},
  journal = {arXiv preprint arXiv:2407.02646},
  year    = {2024},
  url     = {https://arxiv.org/abs/2407.02646}
}

@article{li2025privileged,
  title   = {Do Natural Language Descriptions of Model Activations Convey Privileged Information?},
  author  = {Li, Millicent and Ceballos Arroyo, Alberto Mario and Rogers, Giordano and Saphra, Naomi and Wallace, Byron C.},
  journal = {arXiv preprint arXiv:2509.13316},
  year    = {2025},
  url     = {https://arxiv.org/abs/2509.13316}
}

@inproceedings{ghandeharioun2024patchscopes,
  title     = {Patchscopes: A Unifying Framework for Inspecting Hidden Representations of Language Models},
  author    = {Ghandeharioun, Asma and Caciularu, Avi and Pearce, Adam and Dixon, Lucas and Geva, Mor},
  booktitle = {Proceedings of the 41st International Conference on Machine Learning (ICML)},
  year      = {2024},
  url       = {https://arxiv.org/abs/2401.06102}
}

@article{pan2024latentqa,
  title   = {LatentQA: Teaching LLMs to Decode Activations Into Natural Language},
  author  = {Pan, Alexander and Chen, Lijie and Steinhardt, Jacob},
  journal = {arXiv preprint arXiv:2412.08686},
  year    = {2024},
  url     = {https://arxiv.org/abs/2412.08686}
}

@inproceedings{cohenwang2024contextcite,
  title     = {{ContextCite}: Attributing Model Generation to Context},
  author    = {Cohen-Wang, Benjamin and Shah, Harshay and Georgiev, Kristian and M{\k{a}}dry, Aleksander},
  booktitle = {Advances in Neural Information Processing Systems (NeurIPS)},
  volume    = {37},
  year      = {2024},
  doi       = {10.52202/079017-3035},
  url       = {https://proceedings.neurips.cc/paper_files/paper/2024/hash/adbea136219b64db96a9941e4249a857-Abstract.html}
}

@article{latif2023sparks,
  author  = {Latif, Siddique and Shoukat, Moazzam and Shamshad, Fahad and
             Usama, Muhammad and Ren, Yi and Cuay{\'a}huitl, Heriberto and
             Wang, Wenwu and Zhang, Xulong and Togneri, Roberto and
             Cambria, Erik and Schuller, Bj{\"o}rn W.},
  title   = {Sparks of Large Audio Models: A Survey and Outlook},
  journal = {arXiv preprint arXiv:2308.12792},
  year    = {2023},
}

@misc{runway2026solaris,
  author       = {{Runway}},
  title        = {Introducing {Solaris}: An Interface World Model},
  howpublished = {Runway research release},
  month        = aug,
  year         = {2026},
  note         = {Accessed 2026-09-06},
}

@inproceedings{schemmer2022metaanalysis,
  author    = {Schemmer, Max and Hemmer, Patrick and Nitsche, Maximilian and K{\"u}hl, Niklas and V{\"o}ssing, Michael},
  title     = {A Meta-Analysis of the Utility of Explainable Artificial Intelligence in Human-{AI} Decision-Making},
  booktitle = {Proceedings of the 2022 AAAI/ACM Conference on AI, Ethics, and Society (AIES '22)},
  year      = {2022},
  pages     = {617--626},
  publisher = {ACM},
  address   = {New York, NY, USA},
  doi       = {10.1145/3514094.3534128}
}

@inproceedings{bansal2021whole,
  author    = {Bansal, Gagan and Wu, Tongshuang and Zhou, Joyce and Fok, Raymond and Nushi, Besmira and Kamar, Ece and Ribeiro, Marco Tulio and Weld, Daniel S.},
  title     = {Does the Whole Exceed Its Parts? {T}he Effect of {AI} Explanations on Complementary Team Performance},
  booktitle = {Proceedings of the 2021 CHI Conference on Human Factors in Computing Systems (CHI '21)},
  year      = {2021},
  articleno = {81},
  numpages  = {16},
  publisher = {ACM},
  address   = {New York, NY, USA},
  doi       = {10.1145/3411764.3445717}
}

@article{10.1145/3449287,
author = {Bu{\c c}inca, Zana and Malaya, Maja Barbara and Gajos, Krzysztof Z.},
title = {To Trust or to Think: Cognitive Forcing Functions Can Reduce Overreliance on AI in AI-assisted Decision-making},
year = {2021},
issue_date = {April 2021},
publisher = {Association for Computing Machinery},
address = {New York, NY, USA},
volume = {5},
number = {CSCW1},
url = {https://doi.org/10.1145/3449287},
doi = {10.1145/3449287},
journal = {Proc. ACM Hum.-Comput. Interact.},
month = apr,
articleno = {188},
numpages = {21}
}

@article{wang2025uitars2,
  title   = {{UI-TARS-2} Technical Report: Advancing {GUI} Agent with Multi-Turn Reinforcement Learning},
  author  = {Wang, Haoming and others},
  journal = {arXiv preprint arXiv:2509.02544},
  year    = {2025},
  url     = {https://arxiv.org/abs/2509.02544}
}

@inproceedings{abercrombie-etal-2023-mirages,
  title     = {Mirages. On Anthropomorphism in Dialogue Systems},
  author    = {Abercrombie, Gavin and Cercas Curry, Amanda and Dinkar, Tanvi and Rieser, Verena and Talat, Zeerak},
  editor    = {Bouamor, Houda and Pino, Juan and Bali, Kalika},
  booktitle = {Proceedings of the 2023 Conference on Empirical Methods in Natural Language Processing},
  month     = dec,
  year      = {2023},
  address   = {Singapore},
  publisher = {Association for Computational Linguistics},
  pages     = {4776--4790},
  doi       = {10.18653/v1/2023.emnlp-main.290},
  url       = {https://aclanthology.org/2023.emnlp-main.290}
}

@misc{parkerholder2025genie3,
  title        = {{Genie 3}: A New Frontier for World Models},
  author       = {Parker-Holder, Jack and Fruchter, Shlomi and others},
  year         = {2025},
  howpublished = {Google DeepMind. \url{https://deepmind.google/discover/blog/genie-3-a-new-frontier-for-world-models/}}
}

@article{bai2025qwen3vl,
  title   = {{Qwen3-VL} Technical Report},
  author  = {Bai, Shuai and Cai, Yuxuan and Chen, Ruizhe and Chen, Keqin and Chen, Xionghui and Cheng, Zesen and Deng, Lianghao and Ding, Wei and Gao, Chang and Ge, Chunjiang and others},
  journal = {arXiv preprint arXiv:2511.21631},
  year    = {2025},
  url     = {https://arxiv.org/abs/2511.21631}
}

@inproceedings{miller2023evaluative,
  author    = {Miller, Tim},
  title     = {Explainable {AI} is Dead, Long Live Explainable {AI}! Hypothesis-driven Decision Support using Evaluative {AI}},
  booktitle = {Proceedings of the 2023 ACM Conference on Fairness, Accountability, and Transparency},
  series    = {FAccT '23},
  year      = {2023},
  pages     = {333--342},
  publisher = {Association for Computing Machinery},
  address   = {New York, NY, USA},
  doi       = {10.1145/3593013.3594001},
  url       = {https://doi.org/10.1145/3593013.3594001}
}

@inproceedings{danry2023askme,
  author    = {Danry, Valdemar and Pataranutaporn, Pat and Mao, Yaoli and Maes, Pattie},
  title     = {Don't Just Tell Me, Ask Me: {AI} Systems that Intelligently Frame Explanations as Questions Improve Human Logical Discernment Accuracy over Causal {AI} Explanations},
  booktitle = {Proceedings of the 2023 CHI Conference on Human Factors in Computing Systems},
  series    = {CHI '23},
  year      = {2023},
  articleno = {352},
  numpages  = {13},
  publisher = {Association for Computing Machinery},
  address   = {New York, NY, USA},
  doi       = {10.1145/3544548.3580672},
  url       = {https://doi.org/10.1145/3544548.3580672}
}

@article{kenny2026cwnet,
  author    = {Kenny, Eoin M. and Dharmavaram, Akshay and Lee, Sang Uk and Phan-Minh, Tung and Rajesh, Shreyas and Hu, Yunqing and Major, Laura and Tomov, Momchil S. and Shah, Julie A.},
  title     = {Explainable deep learning improves human mental models of self-driving cars},
  journal   = {Nature},
  volume    = {657},
  number    = {8130},
  pages     = {114--120},
  year      = {2026},
  month     = sep,
  publisher = {Nature Publishing Group},
  doi       = {10.1038/s41586-026-10950-5},
  url       = {https://doi.org/10.1038/s41586-026-10950-5}
}

@article{steyvers2025calibration,
  author    = {Steyvers, Mark and Tejeda, Heliodoro and Kumar, Aakriti and Belem, Catarina and Karny, Sheer and Hu, Xinyue and Mayer, Lukas W. and Smyth, Padhraic},
  title     = {What large language models know and what people think they know},
  journal   = {Nature Machine Intelligence},
  volume    = {7},
  number    = {2},
  pages     = {221--231},
  year      = {2025},
  month     = feb,
  publisher = {Nature Publishing Group},
  doi       = {10.1038/s42256-024-00976-7},
  url       = {https://doi.org/10.1038/s42256-024-00976-7}
}

@article{chen2023intuition,
  author    = {Chen, Valerie and Liao, Q. Vera and Vaughan, Jennifer Wortman and Bansal, Gagan},
  title     = {Understanding the Role of Human Intuition on Reliance in Human-{AI} Decision-Making with Explanations},
  journal   = {Proceedings of the ACM on Human-Computer Interaction},
  volume    = {7},
  number    = {CSCW2},
  articleno = {370},
  numpages  = {32},
  year      = {2023},
  month     = oct,
  publisher = {Association for Computing Machinery},
  address   = {New York, NY, USA},
  doi       = {10.1145/3610219},
  url       = {https://doi.org/10.1145/3610219}
}

@article{vered2023explanations,
  author  = {Vered, Mor and Livni, Tali and Howe, Piers Douglas Lionel and Miller, Tim and Sonenberg, Liz},
  title   = {The Effects of Explanations on Automation Bias},
  journal = {Artificial Intelligence},
  volume  = {322},
  pages   = {103952},
  year    = {2023},
  doi     = {10.1016/j.artint.2023.103952}
}

@inproceedings{schemmer2022automationbias,
  author    = {Schemmer, Max and K{\"u}hl, Niklas and Benz, Carina and Satzger, Gerhard},
  title     = {On the Influence of Explainable {AI} on Automation Bias},
  booktitle = {Proceedings of the 30th European Conference on Information Systems (ECIS 2022)},
  year      = {2022},
  address   = {Timisoara, Romania}
}

@article{lam2023graphcast,
  author  = {Lam, Remi and Sanchez-Gonzalez, Alvaro and Willson, Matthew and
             Wirnsberger, Peter and Fortunato, Meire and Alet, Ferran and
             Ravuri, Suman and Ewalds, Timo and Eaton-Rosen, Zach and
             Hu, Weihua and Merose, Alexander and Hoyer, Stephan and
             Holland, George and Vinyals, Oriol and Stott, Jacklynn and
             Pritzel, Alexander and Mohamed, Shakir and Battaglia, Peter},
  title   = {Learning Skillful Medium-Range Global Weather Forecasting},
  journal = {Science},
  volume  = {382},
  number  = {6677},
  pages   = {1416--1421},
  year    = {2023},
  doi     = {10.1126/science.adi2336},
}

@inproceedings{liang2024foundation,
  author    = {Liang, Yuxuan and Wen, Haomin and Nie, Yuqi and Jiang, Yushan
               and Jin, Ming and Song, Dongjin and Pan, Shirui and Wen, Qingsong},
  title     = {Foundation Models for Time Series Analysis: A Tutorial and Survey},
  booktitle = {Proceedings of the 30th ACM SIGKDD Conference on Knowledge
               Discovery and Data Mining (KDD)},
  pages     = {6555--6565},
  year      = {2024},
  doi       = {10.1145/3637528.3671451},
}

@misc{deepmind2025alphafold,
  author       = {{Google DeepMind}},
  title        = {{AlphaFold}: Five Years of Impact},
  howpublished = {Google DeepMind Blog},
  year         = {2025},
  note         = {Accessed 2026-09-06},
}

@inproceedings{saphra2024mechanistic,
  author    = {Saphra, Naomi and Wiegreffe, Sarah},
  title     = {Mechanistic?},
  booktitle = {Proceedings of the 7th BlackboxNLP Workshop: Analyzing and Interpreting Neural Networks for NLP},
  publisher = {Association for Computational Linguistics},
  address   = {Miami, Florida, USA},
  pages     = {480--498},
  year      = {2024},
  doi       = {10.18653/v1/2024.blackboxnlp-1.30}
}

@inproceedings{sarti2023inseq,
  title     = {Inseq: An Interpretability Toolkit for Sequence Generation Models},
  author    = {Sarti, Gabriele and Feldhus, Nils and Sickert, Ludwig and van der Wal, Oskar},
  editor    = {Bollegala, Danushka and Huang, Ruihong and Ritter, Alan},
  booktitle = {Proceedings of the 61st Annual Meeting of the Association for Computational Linguistics (Volume 3: System Demonstrations)},
  month     = jul,
  year      = {2023},
  address   = {Toronto, Canada},
  publisher = {Association for Computational Linguistics},
  pages     = {421--435},
  doi       = {10.18653/v1/2023.acl-demo.40},
  url       = {https://aclanthology.org/2023.acl-demo.40/}
}

@inproceedings{chen2024selfie,
  title     = {SelfIE: Self-Interpretation of Large Language Model Embeddings},
  author    = {Chen, Haozhe and Vondrick, Carl and Mao, Chengzhi},
  booktitle = {Proceedings of the 41st International Conference on Machine Learning (ICML)},
  year      = {2024},
  url       = {https://arxiv.org/abs/2403.10949}
}

@article{schut2025bridging,
  title   = {Bridging the human--{AI} knowledge gap through concept discovery and transfer in {AlphaZero}},
  author  = {Schut, Lisa and Toma{\v{s}}ev, Nenad and McGrath, Thomas and Hassabis, Demis and Paquet, Ulrich and Kim, Been},
  journal = {Proceedings of the National Academy of Sciences},
  volume  = {122},
  number  = {13},
  pages   = {e2406675122},
  year    = {2025},
  doi     = {10.1073/pnas.2406675122},
  note    = {Preprint: arXiv:2310.16410 (2023)}
}

@article{mcgrath2022acquisition,
  title   = {Acquisition of chess knowledge in {AlphaZero}},
  author  = {McGrath, Thomas and Kapishnikov, Andrei and Toma{\v{s}}ev, Nenad and Pearce, Adam and Wattenberg, Martin and Hassabis, Demis and Kim, Been and Paquet, Ulrich and Kramnik, Vladimir},
  journal = {Proceedings of the National Academy of Sciences},
  volume  = {119},
  number  = {47},
  pages   = {e2206625119},
  year    = {2022},
  doi     = {10.1073/pnas.2206625119},
  note    = {Preprint: arXiv:2111.09259 (2021)}
}

@article{shin2023superhuman,
  title   = {Superhuman artificial intelligence can improve human decision-making by increasing novelty},
  author  = {Shin, Minkyu and Kim, Jin and van Opheusden, Bas and Griffiths, Thomas L.},
  journal = {Proceedings of the National Academy of Sciences},
  volume  = {120},
  number  = {12},
  pages   = {e2214840120},
  year    = {2023},
  doi     = {10.1073/pnas.2214840120},
  note    = {Preprint: arXiv:2303.07462}
}

@article{romeraparedes2024funsearch,
  title   = {Mathematical discoveries from program search with large language models},
  author  = {Romera-Paredes, Bernardino and Barekatain, Mohammadamin and Novikov, Alexander and Balog, Matej and Kumar, M. Pawan and Dupont, Emilien and Ruiz, Francisco J. R. and Ellenberg, Jordan S. and Wang, Pengming and Fawzi, Omar and Kohli, Pushmeet and Fawzi, Alhussein},
  journal = {Nature},
  volume  = {625},
  number  = {7995},
  pages   = {468--475},
  year    = {2024},
  doi     = {10.1038/s41586-023-06924-6},
  note    = {Published online 14 December 2023}
}

@article{fawzi2022alphatensor,
  title   = {Discovering faster matrix multiplication algorithms with reinforcement learning},
  author  = {Fawzi, Alhussein and Balog, Matej and Huang, Aja and Hubert, Thomas and Romera-Paredes, Bernardino and Barekatain, Mohammadamin and Novikov, Alexander and Ruiz, Francisco J. R. and Schrittwieser, Julian and Swirszcz, Grzegorz and Silver, David and Hassabis, Demis and Kohli, Pushmeet},
  journal = {Nature},
  volume  = {610},
  number  = {7930},
  pages   = {47--53},
  year    = {2022},
  doi     = {10.1038/s41586-022-05172-4}
}

@inproceedings{meng2022rome,
  title     = {Locating and Editing Factual Associations in {GPT}},
  author    = {Meng, Kevin and Bau, David and Andonian, Alex and Belinkov, Yonatan},
  booktitle = {Advances in Neural Information Processing Systems},
  volume    = {35},
  year      = {2022},
  eprint    = {2202.05262},
  archivePrefix = {arXiv},
  primaryClass  = {cs.CL},
  url       = {https://arxiv.org/abs/2202.05262},
  note      = {Code and data at \url{https://rome.baulab.info/}}
}

@misc{lu2026assistantaxis,
  title        = {The Assistant Axis: Situating and Stabilizing the Default Persona of Language Models},
  author       = {Christina Lu and Jack Gallagher and Jonathan Michala and Kyle Fish and Jack Lindsey},
  year         = {2026},
  eprint       = {2601.10387},
  archivePrefix = {arXiv},
  primaryClass = {cs.CL},
  url          = {https://arxiv.org/abs/2601.10387}
}

@misc{mccoy2026emergent,
  author        = {McCoy, R. Thomas and Soulos, Paul and Linzen, Tal and Smolensky, Paul},
  title         = {The Emergent Symbolic Structure of Artificial Neural Networks},
  year          = {2026},
  eprint        = {2608.29530},
  archivePrefix = {arXiv},
  primaryClass  = {cs.CL},
  url           = {https://arxiv.org/abs/2608.29530}
}

@article{frasertaliente2026nla,
  title   = {Natural Language Autoencoders Produce Unsupervised Explanations of LLM Activations},
  author  = {Fraser-Taliente, Kit and Kantamneni, Subhash and Ong, Euan and Mossing, Dan and Lu, Christina and Bogdan, Paul C. and Ameisen, Emmanuel and Chen, James and Kishylau, Dzmitry and Pearce, Adam and Tarng, Julius and Wu, Alex and Wu, Jeff and Zhang, Yang and Ziegler, Daniel M. and Hubinger, Evan and Batson, Joshua and Lindsey, Jack and Zimmerman, Samuel and Marks, Samuel},
  journal = {Transformer Circuits Thread},
  year    = {2026},
  url     = {https://transformer-circuits.pub/2026/nla/index.html}
}

@article{karvonen2025activation,
  title   = {Activation Oracles: Training and Evaluating LLMs as General-Purpose Activation Explainers},
  author  = {Karvonen, Adam and Chua, James and Dumas, Cl{\'e}ment and Fraser-Taliente, Kit and Kantamneni, Subhash and Minder, Julian and Ong, Euan and Sen Sharma, Arnab and Wen, Daniel and Evans, Owain and Marks, Samuel},
  journal = {arXiv preprint arXiv:2512.15674},
  year    = {2025},
  url     = {https://arxiv.org/abs/2512.15674}
}

@article{marks2025auditing,
  title   = {Auditing Language Models for Hidden Objectives},
  author  = {Marks, Samuel and Treutlein, Johannes and Bricken, Trenton and Lindsey, Jack and Marcus, Jonathan and Mishra-Sharma, Siddharth and Ziegler, Daniel and Ameisen, Emmanuel and Batson, Joshua and Belonax, Tim and Bowman, Samuel R. and Carter, Shan and Chen, Brian and Cunningham, Hoagy and Denison, Carson and Dietz, Florian and Golechha, Satvik and Khan, Akbir and Kirchner, Jan and Leike, Jan and Meek, Austin and Nishimura-Gasparian, Kei and Ong, Euan and Olah, Christopher and Pearce, Adam and Roger, Fabien and Salle, Jeanne and Shih, Andy and Tong, Meg and Thomas, Drake and Rivoire, Kelley and Jermyn, Adam and MacDiarmid, Monte and Henighan, Tom and Hubinger, Evan},
  journal = {arXiv preprint arXiv:2503.10965},
  year    = {2025},
  url     = {https://arxiv.org/abs/2503.10965}
}

@article{sharkey2025open,
  title   = {Open Problems in Mechanistic Interpretability},
  author  = {Sharkey, Lee and Chughtai, Bilal and Batson, Joshua and Lindsey, Jack and Wu, Jeff and Bushnaq, Lucius and Goldowsky-Dill, Nicholas and Heimersheim, Stefan and Ortega, Alejandro and Bloom, Joseph and Biderman, Stella and Garriga-Alonso, Adria and Conmy, Arthur and Nanda, Neel and Rumbelow, Jessica and Wattenberg, Martin and Schoots, Nandi and Miller, Joseph and Michaud, Eric J. and Casper, Stephen and Tegmark, Max and Saunders, William and Bau, David and Todd, Eric and Geiger, Atticus and Geva, Mor and Hoogland, Jesse and Murfet, Daniel and McGrath, Tom},
  journal = {arXiv preprint arXiv:2501.16496},
  year    = {2025},
  url     = {https://arxiv.org/abs/2501.16496}
}

@inproceedings{fitzmaurice1995tangiblebits,
  author    = {Fitzmaurice, George W. and Ishii, Hiroshi and Buxton, William A. S.},
  title     = {Bricks: Laying the Foundations for Graspable User Interfaces},
  booktitle = {Proceedings of the SIGCHI Conference on Human Factors in Computing Systems (CHI '95)},
  year      = {1995},
  pages     = {442--449},
  publisher = {ACM Press/Addison-Wesley},
  address   = {New York, NY, USA},
  doi       = {10.1145/223904.223964}
}

@inproceedings{pataranutaporn2020livingbits,
  author    = {Pataranutaporn, Pat and Vujic, Angela and Kong, David S. and Maes, Pattie and Sra, Misha},
  title     = {Living Bits: Opportunities and Challenges for Integrating Living Microorganisms in Human-Computer Interaction},
  booktitle = {Proceedings of the Augmented Humans International Conference (AHs '20)},
  year      = {2020},
  articleno = {30},
  pages     = {1--12},
  publisher = {ACM},
  address   = {New York, NY, USA},
  doi       = {10.1145/3384657.3384783}
}

@inproceedings{bhatt2021uncertainty,
  title     = {Uncertainty as a Form of Transparency: Measuring,
               Communicating, and Using Uncertainty},
  author    = {Bhatt, Umang and Antor{\'a}n, Javier and Zhang, Yunfeng and
               Liao, Q. Vera and Sattigeri, Prasanna and Fogliato, Riccardo
               and Melan{\c{c}}on, Gabrielle and Krishnan, Ranganath and
               Stanley, Jason and Tickoo, Omesh and Nachman, Lama and
               Chunara, Rumi and Srikumar, Madhulika and Weller, Adrian and
               Xiang, Alice},
  booktitle = {Proceedings of the 2021 AAAI/ACM Conference on AI, Ethics,
               and Society (AIES)},
  pages     = {401--413},
  year      = {2021},
  doi       = {10.1145/3461702.3462571},
  url       = {https://arxiv.org/abs/2011.07586},
  note      = {Communicating uncertainty (not just point estimates) aids
               user decisions; verify pages against final version}
}

@inproceedings{ding2025citations,
  title     = {Citations and Trust in {LLM} Generated Responses},
  author    = {Ding, Yifan and Facciani, Matthew and Poudel, Amrit and
               Joyce, Ellen and Aguinaga, Salvador and Veeramani, Balaji and
               Bhattacharya, Sanmitra and Weninger, Tim},
  booktitle = {Proceedings of the AAAI Conference on Artificial
               Intelligence},
  volume    = {39},
  number    = {22},
  pages     = {23787--23795},
  year      = {2025},
  doi       = {10.1609/aaai.v39i22.34550},
  url       = {https://arxiv.org/abs/2501.01303},
  note      = {Citations raise self-reported trust even when random;
               trust drops when users check them}
}

@article{lee2004trust,
  title   = {Trust in Automation: Designing for Appropriate Reliance},
  author  = {Lee, John D. and See, Katrina A.},
  journal = {Human Factors},
  volume  = {46},
  number  = {1},
  pages   = {50--80},
  year    = {2004},
  doi     = {10.1518/hfes.46.1.50_30392},
}

@inproceedings{ehsan2024seamful,
  title     = {Seamful {XAI}: Operationalizing Seamful Design in
               Explainable {AI}},
  author    = {Ehsan, Upol and Liao, Q. Vera and Passi, Samir and
               Riedl, Mark O. and Daum{\'e} III, Hal},
  booktitle = {Proceedings of the 2024 CHI Conference on Human Factors in
               Computing Systems},
  year      = {2024},
  doi       = {10.1145/3613904.3642755},
  url       = {https://arxiv.org/abs/2211.06753},
}

@inproceedings{kizilcec2016howmuchinfo,
  author    = {Kizilcec, Ren{\'e} F.},
  title     = {How Much Information? Effects of Transparency on Trust in an Algorithmic Interface},
  booktitle = {Proceedings of the 2016 CHI Conference on Human Factors in Computing Systems (CHI '16)},
  pages     = {2390--2395},
  publisher = {Association for Computing Machinery},
  address   = {New York, NY, USA},
  year      = {2016},
  doi       = {10.1145/2858036.2858402}
}

@inproceedings{springer2019progressive,
  author    = {Springer, Aaron and Whittaker, Steve},
  title     = {Progressive Disclosure: Empirically Motivated Approaches to Designing Effective Transparency},
  booktitle = {Proceedings of the 24th International Conference on Intelligent User Interfaces (IUI '19)},
  pages     = {107--120},
  publisher = {Association for Computing Machinery},
  address   = {New York, NY, USA},
  year      = {2019},
  doi       = {10.1145/3301275.3302322}
}

@article{springer2020disclosure,
author = {Springer, Aaron and Whittaker, Steve},
title = {Progressive Disclosure: When, Why, and How Do Users Want Algorithmic Transparency Information?},
year = {2020},
issue_date = {December 2020},
publisher = {Association for Computing Machinery},
address = {New York, NY, USA},
volume = {10},
number = {4},
issn = {2160-6455},
url = {https://doi.org/10.1145/3374218},
doi = {10.1145/3374218},
journal = {ACM Trans. Interact. Intell. Syst.},
month = oct,
articleno = {29},
numpages = {32}
}

@article{zaosanders2025genai,
  title   = {How People Are Really Using Gen {AI} in 2025},
  author  = {Zao-Sanders, Marc},
  journal = {Harvard Business Review},
  year    = {2025},
  month   = apr,
  day     = {9},
  url     = {https://hbr.org/2025/04/how-people-are-really-using-gen-ai-in-2025},
  note    = {Practitioner analysis of public forum discourse; ranks
             therapy/companionship as the top use case. Not peer-reviewed}
}

@article{zhang2026companions,
  title   = {Interaction with {AI} companions and psychological well-being},
  author  = {Zhang, Yutong and Zhao, Dora and Hancock, Jeffrey T. and
             Kraut, Robert and Yang, Diyi},
  journal = {Nature Human Behaviour},
  year    = {2026},
  doi     = {10.1038/s41562-026-02516-2},
  url     = {https://www.nature.com/articles/s41562-026-02516-2},
  note    = {Mixed-methods study of 1,131 Character.AI users; asks users
             their primary motivation (productivity, entertainment,
             curiosity, relational). Preprint: arXiv:2506.12605}
}

@article{miller2026unconditional,
  title   = {Artificial intelligence and unconditional love: The rise of
             generative {AI} as an alternative form of mental health support},
  author  = {Miller, Vince and Hill, Mark J. and Moreira, Tiago},
  journal = {Communication and Change},
  volume  = {2},
  pages   = {Article 14},
  year    = {2026},
  doi     = {10.1007/s44382-026-00036-1},
  url     = {https://link.springer.com/article/10.1007/s44382-026-00036-1},
  note    = {Interviews + Reddit analysis; use blends emotional support,
             advice-seeking, entertainment -- closer to companionship
             than task-specific tool use}
}

@article{rousmaniere2025llm,
  title   = {Large Language Models as Mental Health Resources: Patterns of Use in the {United States}},
  author  = {Rousmaniere, Tony and Zhang, Y. and Li, X. and Shah, S.},
  journal = {Practice Innovations},
  volume  = {11},
  number  = {2},
  pages   = {139--155},
  year    = {2026},
  month   = jun,
  doi     = {10.1037/pri0000292}
}

@misc{lesswrongInterpretingGPT,
	author = {nostalgebraist},
	title = {interpreting {G}{P}{T}: the logit lens — {L}ess{W}rong --- lesswrong.com},
	howpublished = {\url{https://www.lesswrong.com/posts/AcKRB8wDpdaN6v6ru/interpreting-gpt-the-logit-lens}},
	year = {2020},
	note = {[Accessed 12-09-2025]},
}

@article{lieberum2024gemmascope,
  title        = {Gemma Scope: Open Sparse Autoencoders Everywhere All At Once on Gemma 2},
  author       = {Lieberum, Tom and Rajamanoharan, Senthooran and Conmy, Arthur and
                  Smith, Lewis and Sonnerat, Nicolas and Varma, Vikrant and
                  Kram{\'a}r, J{\'a}nos and Dragan, Anca and Shah, Rohin and Nanda, Neel},
  journal      = {arXiv preprint arXiv:2408.05147},
  year         = {2024}
}

@book{zwicky1969discovery,
  author    = {Zwicky, Fritz},
  title     = {Discovery, Invention, Research through the Morphological Approach},
  publisher = {The Macmillan Company},
  address   = {Toronto},
  year      = {1969}
}

@article{ritchey2018gma,
  author  = {Ritchey, Tom},
  title   = {General Morphological Analysis as a Basic Scientific Modelling Method},
  journal = {Technological Forecasting and Social Change},
  volume  = {126},
  pages   = {81--91},
  year    = {2018},
  doi     = {10.1016/j.techfore.2017.05.027}
}

@article{bergen2026anatomy,
  title         = {Anatomy of Post-Training: Using Interpretability to Characterize
                   Data and Shape the Learning Signal},
  author        = {Bergen, Leon and Bhalla, Usha and Baskaran, Sidharth and
                   Loeffler, Max and Sarfati, Raphael and Gala, Dhruvil and
                   Panwar, Ryan and Aranguri, Santiago and Fel, Thomas and
                   Geiger, Atticus and Kowal, Matthew and Boppana, Siddharth and
                   Balsam, Daniel and Lewis, Owen and Merullo, Jack and
                   McGrath, Thomas and Lubana, Ekdeep Singh},
  journal       = {arXiv preprint arXiv:2606.12360},
  year          = {2026},
  eprint        = {2606.12360},
  archivePrefix = {arXiv},
  primaryClass  = {cs.LG}
}

@article{paek2025audiolatent,
  title        = {Learning Interpretable Features in Audio Latent Spaces via Sparse Autoencoders},
  author       = {Paek, Nathan and Zang, Yongyi and Yang, Qihui and Leistikow, Randal},
  journal      = {arXiv preprint arXiv:2510.23802},
  year         = {2025}
}

@inproceedings{aparin2026audiosae,
  title        = {{AudioSAE}: Towards Understanding of Audio-Processing Models with Sparse {AutoEncoders}},
  author       = {Aparin, Georgii and Sadekova, Tasnima and Rukhovich, Alexey and Yermekova, Assel and Kushnareva, Laida and Popov, Vadim and Kuznetsov, Kristian and Piontkovskaya, Irina},
  booktitle    = {Proceedings of the 19th Conference of the European Chapter of the Association for Computational Linguistics (Volume 1: Long Papers)},
  pages        = {3221--3254},
  year         = {2026},
  address      = {Rabat, Morocco},
  publisher = {Association for Computational Linguistics},
}

@misc{lindsey2024crosscoders,
  title={{Sparse Crosscoders for Cross-Layer Features and Model Diffing}},
  author={Lindsey, Jack and Templeton, Adly and Marcus, Jonathan and Conerly, Thomas and Batson, Joshua and Olah, Christopher},
  howpublished={Transformer Circuits Thread},
  note={\url{https://transformer-circuits.pub/2024/crosscoders/index.html}},
  year={2024}
}

@article{wesp2026medical,
  title={{Sparse Autoencoders for Interpretable Medical Image Representation Learning}},
  author={Wesp, Philipp and Holland, Robbie and Sideri-Lampretsa, Vasiliki and Gatidis, Sergios},
  journal={arXiv preprint arXiv:2603.23794},
  year={2026}
}

@article{kim2025conceptsteersparse,
  title={{Concept Steerers: Leveraging K-Sparse Autoencoders for Test-Time Controllable Generations}},
  author={Kim, Dahye and Ghadiyaram, Deepti},
  journal={arXiv preprint arXiv:2501.19066},
  year={2025}
}

@inproceedings{cywinski2025saeuron,
  title={{SAeUron: Interpretable Concept Unlearning in Diffusion Models with Sparse Autoencoders}},
  author={Cywi{\'n}ski, Bartosz and Deja, Kamil},
  booktitle={Proceedings of the 42nd International Conference on Machine Learning (ICML)},
  year={2025}
}

@article{joseph2025steeringsae,
  title={{Steering CLIP's Vision Transformer with Sparse Autoencoders}},
  author={Joseph, Sonia and Suresh, Praneet and Goldfarb, Ethan and Hufe, Lorenz and Gandelsman, Yossi and Graham, Robert and Bzdok, Danilo and Samek, Wojciech and Richards, Blake Aaron},
  journal={arXiv preprint arXiv:2504.08729},
  year={2025}
}

@inproceedings{zhang2023addingcontrolsae,
  title={{Adding Conditional Control to Text-to-Image Diffusion Models}},
  author={Zhang, Lvmin and Rao, Anyi and Agrawala, Maneesh},
  booktitle={Proceedings of the IEEE/CVF International Conference on Computer Vision (ICCV)},
  year={2023}
}

@inproceedings{gandikota2024conceptsae,
  title={{Concept Sliders: LoRA Adaptors for Precise Control in Diffusion Models}},
  author={Gandikota, Rohit and Materzy{\'n}ska, Joanna and Zhou, Tingrui and Torralba, Antonio and Bau, David},
  booktitle={European Conference on Computer Vision (ECCV)},
  year={2024}
}

@article{marks2023geometry,
  title={The geometry of truth: Emergent linear structure in large language model representations of true/false datasets},
  author={Marks, Samuel and Tegmark, Max},
  journal={arXiv preprint arXiv:2310.06824},
  year={2023}
}

@article{kim2025linear,
  title={Linear representations of political perspective emerge in large language models},
  author={Kim, Junsol and Evans, James and Schein, Aaron},
  journal={arXiv preprint arXiv:2503.02080},
  year={2025}
}

@article{lindsey2025biology,
  author={Lindsey, Jack and Gurnee, Wes and Ameisen, Emmanuel and Chen, Brian and Pearce, Adam and Turner, Nicholas L. and Citro, Craig and Abrahams, David and Carter, Shan and Hosmer, Basil and Marcus, Jonathan and Sklar, Michael and Templeton, Adly and Bricken, Trenton and McDougall, Callum and Cunningham, Hoagy and Henighan, Thomas and Jermyn, Adam and Jones, Andy and Persic, Andrew and Qi, Zhenyi and Thompson, T. Ben and Zimmerman, Sam and Rivoire, Kelley and Conerly, Thomas and Olah, Chris and Batson, Joshua},
  title={On the Biology of a Large Language Model},
  journal={Transformer Circuits Thread},
  year={2025},
  url={https://transformer-circuits.pub/2025/attribution-graphs/biology.html}
}

@article{turner2023steering,
  title={Steering language models with activation engineering},
  author={Turner, Alexander Matt and Thiergart, Lisa and Leech, Gavin and Udell, David and Vazquez, Juan J and Mini, Ulisse and MacDiarmid, Monte},
  journal={arXiv preprint arXiv:2308.10248},
  year={2023}
}

@inproceedings{rimsky2024steering,
  title={Steering llama 2 via contrastive activation addition},
  author={Rimsky, Nina and Gabrieli, Nick and Schulz, Julian and Tong, Meg and Hubinger, Evan and Turner, Alexander},
  booktitle={Proceedings of the 62nd Annual Meeting of the Association for Computational Linguistics (Volume 1: Long Papers)},
  pages={15504--15522},
  year={2024}
}

@article{chu2025illusions,
  title={Illusions of Intimacy: How Emotional Dynamics Shape Human-AI Relationships},
  author={Chu, Minh Duc and Gerard, Patrick and Pawar, Kshitij and Bickham, Charles and Lerman, Kristina},
  journal={arXiv preprint arXiv:2505.11649},
  year={2025}
}

@article{cheng2026sycophantic,
  title={Sycophantic AI decreases prosocial intentions and promotes dependence},
  author={Cheng, Myra and Lee, Cinoo and Khadpe, Pranav and Yu, Sunny and Han, Dyllan and Jurafsky, Dan},
  journal={Science},
  volume={391},
  number={6792},
  pages={eaec8352},
  year={2026},
  publisher={American Association for the Advancement of Science}
}

@article{wurgaft2026manifold,
  title={Manifold steering reveals the shared geometry of neural network representation and behavior},
  author={Wurgaft, Daniel and Rager, Can and Kowal, Matthew and Shyam, Vasudev and Feucht, Sheridan and Bhalla, Usha and Haklay, Tal and Bigelow, Eric and Sarfati, Raphael and McGrath, Thomas and others},
  journal={arXiv preprint arXiv:2605.05115},
  year={2026}
}

@inproceedings{karny2026neural,
  title={Neural transparency: Mechanistic interpretability interfaces for anticipating model behaviors for personalized ai},
  author={Karny, Sheer and Baez, Anthony and Pataranutaporn, Pat},
  booktitle={Proceedings of the 31st International Conference on Intelligent User Interfaces},
  pages={868--884},
  year={2026}
}

@article{marcilio2026navigating,
  title={Navigating the Concept Space of Language Models},
  author={Marc{\'\i}lio-Jr, Wilson E and Eler, Danilo M},
  journal={arXiv preprint arXiv:2603.23524},
  year={2026}
}

@article{yan2025visual,
  title={Visual Exploration of Feature Relationships in Sparse Autoencoders with Curated Concepts},
  author={Yan, Xinyuan and Liu, Shusen and Thopalli, Kowshik and Wang, Bei},
  journal={arXiv preprint arXiv:2511.06048},
  year={2025}
}

@article{li2025conceptviz,
  title={ConceptViz: A Visual Analytics Approach for Exploring Concepts in Large Language Models},
  author={Li, Haoxuan and Wen, Zhen and Jiang, Qiqi and Li, Chenxiao and Wu, Yuwei and Yang, Yuchen and Wang, Yiyao and Huang, Xiuqi and Zhu, Minfeng and Chen, Wei},
  journal={IEEE Transactions on Visualization and Computer Graphics},
  year={2025},
  publisher={IEEE}
}

@article{ameisen2025circuit,
	title        = {Circuit Tracing: Revealing Computational Graphs in Language Models},
	author       = {Ameisen, Emmanuel and Lindsey, Jack and Pearce, Adam and Gurnee, Wes and Turner, Nicholas L. and Chen, Brian and Citro, Craig and Abrahams, David and Carter, Shan and Hosmer, Basil and Marcus, Jonathan and Sklar, Michael and Templeton, Adly and Bricken, Trenton and McDougall, Callum and Cunningham, Hoagy and Henighan, Thomas and Jermyn, Adam and Jones, Andy and Persic, Andrew and Qi, Zhenyi and Ben Thompson, T. and Zimmerman, Sam and Rivoire, Kelley and Conerly, Thomas and Olah, Chris and Batson, Joshua},
	year         = 2025,
	journal      = {Transformer Circuits Thread},
	url          = {https://transformer-circuits.pub/2025/attribution-graphs/methods.html}
}

@article{viegas2023system,
	title        = {The system model and the user model: Exploring AI dashboard design},
	author       = {Vi{\'e}gas, Fernanda and Wattenberg, Martin},
	year         = 2023,
	journal      = {arXiv preprint arXiv:2305.02469}
}

\appendix

\end{document}